# Revisiting with *Chandra* the Scaling Relations of the X-ray Emission Components (Binaries, Nuclei and Hot Gas) of Early Type Galaxies

Bram Boroson, Dong-Woo Kim, and Giuseppina Fabbiano
Smithsonian Astrophysical Observatory, 60 Garden Street, Cambridge, MA 02138
(Wednesday, November 10, 2010)

## Abstract

We have selected a sample of 30 normal (non-cD) early type galaxies, for all of which optical spectroscopy is available, and which have been observed with Chandra to a depth such to insure the detection of bright low-mass X-ray binaries (LMXBs) with $L_X > 10^{38}$ erg $s^{-1}$. This sample includes a larger fraction of gas-poor galaxies than previously studied samples, and covers a wide range of stellar luminosity ($L_K$), velocity dispersion ($\sigma_*$), GC specific frequency ($S_N$) and stellar age. We derive X-ray luminosities (or upper limits) from the different significant X-ray components of these galaxies: nuclei, detected and undetected LMXBs, coronally active binaries (ABs), cataclysmic variables (CVs), and hot gas. The ABs and CVs contribution is estimated from the $L_X$-$L_K$ scaling relation of M31 and M32. The contribution of undetected LMXBs is estimated both by fitting the spectra of the unresolved X-ray emission and by extrapolating the LMXB X-ray luminosity function (XLF). On average, the X-ray luminosity of LMXBs is a factor of ~10 higher than that of ABs+CVs. By spectral fitting the emission (also considering gas emission in the regions of point sources) we estimate the contribution of the hot gas. We find our sample equally divided among galaxies with $L_X$(gas) > $L_X$(LMXB), $L_X$(ABCV) ≤ $L_X$(gas) ≤ $L_X$(LMXB) and $L_X$(gas) < $L_X$(ABCV).

The results for the nuclei are consistent with those discussed by Pellegrini (2010). We derive a revised scaling relation between the integrated X-ray luminosity of LMXBs in a galaxy and the $L_K$ luminosity of the host galaxy: $L_X$(LMXB)/$L_K$ ~ $10^{29}$ erg $s^{-1}$ $L_K^{-1}$ with 50% $1\sigma$ rms; moreover, we also obtain a tighter $L_X$(LMXB)/$L_K$ – $S_N$ relation than previously published. We revisit the relations between hot gas content and other galaxy parameters ($L_K$, $\sigma_*$), which in most previous work was based on the integrated total X-ray luminosity of the galaxy, finding a steeper $L_X$(gas)-$L_K$ relation with larger scatter than reported in the literature. We find a positive correlation between the luminosity and temperature of the hot ISM, significantly tighter than reported by earlier studies. This relation is particularly well defined in the subsample with $\sigma_*$>240 km/s, where it may be related to the analogous correlation found in cD galaxies and groups/clusters. However, the gas-poor galaxies with the shallowest potentials ($\sigma_*$ < 200 km/s) also follow this relation, contrary to the expected anti-correlation in a simple outflow/wind scenario. Galaxies with intermediate values of $\sigma_*$ instead tend to have the same kT, while $L_X$(gas) spans a factor of ~20; among these galaxies, we find a moderate, positive correlation between $L_X$(gas) and the average stellar age, possibly suggesting a transition from halo retention to outflow caused by rejuvenated star formation associated with recent mergers.

## I. Introduction

It is now well established that early-type galaxies (E and S0) emit X-rays from a hot ISM and populations of LMXBs (see e.g., reviews by Fabbiano 1989, 2006). The super-massive nuclear black holes of these galaxies may also in some cases contribute to the emission, with sources ranging from radio-loud luminous AGNs (e.g. in 3CR galaxies; Fabbiano et al 1984) to low-luminosity AGNs (e.g., in NGC 1313, see Kim & Fabbiano 2003; and quieter nuclei, e.g., Pellegrini 2010).

Widespread X-ray emission from early-type galaxies was discovered with the Einstein observatory, the first imaging X-ray telescope (e.g., Forman et al. 1979; Trinchieri & Fabbiano 1985; Forman, Jones & Tucker 1985), but the relative contribution of different types of sources to this emission, and the physical state of the hot ISM, has been debated for years (see e.g., above references; Kim et al. 1992; Eskridge, Fabbiano & Kim 1995; Ciotti et al. 1991). With the sub-arcsecond resolution and sensitivity of the Chandra Observatory, we have been able for the first time to resolve individual point-like sources, such as LMXBs and faint nuclei, in these galaxies (see Fabbiano 2006). By subtraction, these observations can be used to set more stringent – and realistic - constraints on the amount of hot ISM present in a galaxy, especially in those galaxies where the output of LMXB populations dominates the X-ray. Likewise, the luminosity of faint nuclear sources can be constrained, to limits compatible with the X-ray luminosity of LMXBs (Pellegrini 2010).

These high resolution data are essential not only to measure the amount of hot ISM is a given galaxy, but also to obtain correct measurements of the properties of this ISM (luminosity, temperature, metal abundances; see e.g. Kim & Fabbiano 2003), particularly for the hot-ISM poor galaxies. For these galaxies, a simple subtraction of all the detected LMXBs is not enough, and one must account for undetected LMXBs and other stellar sources (Kim & Fabbiano 2004; Revnitsev et al 2008). In particular, X-ray fainter stellar sources such as active binaries (ABs) and cataclysmic variables (CVs), which we see in the Milky Way (Heinke et al. 2008), must be present in external galaxies. The integrated contribution from these sources was reported in M32 (Revnivtsev et al. 2008), but they cannot be individually detected even with Chandra. They are often ignored because of their relatively small contribution to the total X-ray luminosity, as first estimated by Pellegrini & Fabbiano (1994). However, their contribution to the unresolved emission of the gas-poor galaxies is not negligible, once most bright LMXBs are excluded.

In this paper, (1) we seek to obtain as accurately as possible measurements of the luminosity and temperature of the hot gaseous ISM for a selected sample of 30 early-type galaxies, by carefully estimating the contribution from individual sources, including LMXBs (detected and undetected), ABs+CVs and nuclei. We will present a discussion of the metal abundance in a separate paper. (2) Then, we revisit the scaling relation between gas luminosity, and other basic galaxy properties such as integrated stellar luminosity $L_K$, velocity dispersion $\sigma_*$, and globular cluster specific frequency $S_N$. These relations have been the basis for much discussion and modeling of the physical evolution of the hot halos in past studies (see Canizares, Fabbiano & Trinchieri 1987; Eskridge et al. 1995a, b; O'Sullivan et al.

2003; Ciotti et al 1991; Kim & Fabbiano 2004; David et al 2006). Although David et al. (2006) investigated gas properties in gas-poor galaxies with Chandra, these authors did not consider the contribution from ABs and CVs. Besides using accurate measurements for the hot gas contributions, our study includes a larger representation of gas-poor galaxies than found in these previous studies.

This paper is organized as follows. In Section 2, we describe our sample selection. In Section 3, fitting the X-ray spectra with proper emission models, we measure the individual emission components, ABs and CVs, nucleus, detected and undetected LMXBs and hot gas. We also measure the contribution from undetected LMXBs by extrapolating the X-ray luminosity function of LMXBs. In section 4, we present various correlations between the X-ray and optical properties and discuss their implications. In Section 5, we summarize our conclusions.

## 2. Sample Properties

We have selected 30 nearby early type galaxies, which were well studied both in X-ray and optical bands. We excluded cD galaxies, which are dominant galaxies in groups and clusters and are associated with extended hot halos confined by the group/cluster potential. Our sample includes both gas-poor (e.g., M32, del Burgo et al. 2001, Coelho, de Oliveira, & Fernandes 2009; NGC 821: Pellegrini et al. 2007a) and gas-rich galaxies (e.g., NGC 4472, NGC 4649, Fabbiano, Kim, & Trinchieri 1992).

For our sample of galaxies optical line indices measurement are available, providing estimates of the velocity dispersion $\sigma_*$ and age. We list the basic properties of the sample galaxies in Table 1, including morphological types (from RC3), $R_{25}$ (from RC3), distances (from Tonry et al. 2001), ages, $\sigma_*$, B mag (from RC3), $M_B$, K mag (from 2MASS via NED), log $L_K$ (assuming $K_\odot$= 3.33 mag) and the GC specific frequency $S_N$. The optically measured ages and $\sigma_*$ are from the literature. When several measurements are available, we take them in order of Thomas et al. (2005), Trager et al. (2000), Terlevich & Forbes (2002), Howell (2005), Gallagher et al. (2008) and McDermid et al. (2006). We take $S_N$ from the literature in order of Peng et al. (2008), Harris & Harris (1999) and Ashman & Zepf (1998). We note that for some galaxies the reported values of $S_N$ vary widely from one measurement to another. For example, for NGC 4526 Peng et al. (2008) measured $S_N$ = 1.09 with the HST ACS data as part of ACS Virgo Cluster Survey, while $S_N$ = 7.4-7.7 in Kissler-Patig (1997), Ashman & Zepf (1998) and Harris & Harris (1999). Because HST results are more reliable in identifying globular clusters and in reducing contaminations than those based on the ground-based observations, we primarily take $S_N$ from Peng et al. (2008). Our sample provides good coverage of optical/IR luminosity ($L_K$ from $10^{9.1}$ to $10^{11.7}$ $L_{K\odot}$), GC specific frequency ($S_N$ = 1-7), and stellar velocity dispersion ($\sigma_*$ = 160 – 300 km $s^{-1}$, reaching the lowest $\sigma_*$ = 72 and 108 km $s^{-1}$ in M32 and NGC 3377 respectively).

All galaxies were targeted in the Chandra ACIS-S observations for exposure times long enough to detect bright LMXBs with $L_X$ > 1-2 × $10^{38}$ erg $s^{-1}$. The X-ray data are taken from the Chandra archive (http://cxc.harvard.edu/cda). We do not use early ACIS observations, which were taken in 1999 with a CCD temperature below -120 C. In Table 2, we list for each galaxy the Chandra observation id, observation date, exposure time (after excluding background flares), the Galactic line of sight $N_H$ taken from the NRAO survey (Dickey & Lockman 1990), and the point source detection limit derived as explained in Section 3.

## 3. X-ray Data Analysis

The ACIS data were uniformly reduced in a similar manner as described in Kim & Fabbiano (2003) with a custom-made pipeline (XPIPE), specifically developed for the Chandra Multiwavelength Project (ChaMP; Kim et al. 2004). We apply *acis_process_events* to properly correct for the time-dependent gain and charge transfer inefficiency (CTI). For observations taken in and after 2006, we apply the revised ACIS contamination model (see http://cxc.harvard.edu/cal/memos/contam_memo.pdf). We generate a light curve to check for background flares and exclude events occurring during flares (see Kim et al. 2004 for more details). For targets with multiple observations, we re-project the individual observations to a common tangent point and combine them by using *merge_all* available in the CIAO contributed package (http://cxc.harvard.edu /ciao/threads/combine/).

The X-ray point sources were detected using CIAO *wavdetect*. We set the significance threshold to be $10^{-6}$, which corresponds approximately to one false source per chip and the exposure threshold to be 10% using an exposure map. The latter was applied to reduce the false detections often found at the chip edge. To measure the X-ray flux and luminosity (in 0.3-8 keV), we take into account the temporal and spatial QE variation (http://cxc.harvard.edu/cal/Acis/Cal_prods/qeDeg/) by calculating the energy conversion factor (ECF = ratio of flux to count rate) for each source in each observation. To calculate the X-ray flux of sources detected in the merged data, we apply an exposure-weighted mean ECF. This will generate a flux as if the entire observations were done in one exposure, but with a variable detector QE as in the real observations.

The response files, rmf (response matrix file) and arf (ancillary reference file), were generated for each source region. For data taken in multiple exposures, to take into account the ACIS response degradation due to the filter contamination, we generate arf per individual observation and then take an exposure-weighted mean by applying dmarfadd (for weighted sum) and dmtcalc (to divide by the number of observations). The background spectra are extracted from the source free region within the same CCD. The spectra were binned to have a minimum of 25 counts per energy bin.

### 3.1 Stellar X-ray Sources (ABs and CVs)

The contribution from unresolved stellar sources to the X-ray emission of elliptical and S0 galaxies was first considered by Pellegrini & Fabbiano (1994) and has been more recently revisited by Revnivtsev et al. (2008). Stellar sources include active binaries (ABs) and cataclysmic variables (CVs). Typically, $L_X$(AB+CV) is only a small fraction of the total X-ray luminosity, and therefore this contribution was usually ignored in the past. However, it becomes an important factor for constraining the small amounts of hot gas in X-ray-faint ellipticals now that with Chandra we can resolve out the contribution of individual LMXBs and nuclear sources. Revnivtsev et al. (2007a,b, 2009) reported that these stellar sources indeed dominate the unresolved X-ray emission in M32 and the Galactic bulge.

In Appendix A, we report in detail our characterization of the X-ray spectra of a population of ABs and CVs, using Chandra observations of M31 and M32. Because of their proximity, all LMXBs can be detected in both galaxies. The X-ray emission of M32 is entirely due to stellar sources (see also Revnivtsev 2007). Instead, the bulge of M31 contains some hot gas (Bogdan & Gilfanov 2008; Liu et al. 2010). We jointly fit the two spectra of M31 and M32 with a combination of APEC and power-law (PL) models and determine the spectral parameters: kT=0.48 (-0.05, +0.07) keV for AP and $\Gamma$=1.76 $\pm$ 0.37 for PL (see Appendix A; errors quoted here and in the rest of this paper are 1σ). We also derive X-ray to K-band luminosity ratios and corresponding errors in various energy ranges. The ratio in 0.3 – 8 keV, $L_X/L_K$ = $9.5^{+2.1}_{-1.1}$ x $10^{27}$ erg $s^{-1}$ $L_{K\odot}$ can be compared directly with that of Revnivtsev et al. (2007a), who considered M32; while consistent within the errors, the ratio we derive is formally lower than that of Revnivtsev et al. (2008), who considered NGC 3379. Using this ratio, we estimate the expected AB+CV contribution for each galaxy, based on the K-band luminosity ($L_K$) for a given region. To measure the K band magnitude within the source extraction region, we use K band images obtained from the 2MASS Large Galaxy Atlas (Jarrett et al. 2003) whenever available, or the 2MASS All Sky Survey (Skrutskie et al. 2006). We follow absolute photometric calibration of 2MASS discussed by Cohen, Wheaton, & Megeath (2003) and eliminate K band point sources. The resulting $L_X$(AB+CV) is listed in Table 3 for three regions per galaxy (regions for the nucleus, detected LMXBs and the remaining diffuse emission).

## 3.2 Nuclei

To identify the X-ray source at the galactic center and to effectively separate LMXBs near the center, we visually inspect all the Chandra images of individual galaxies. We use the 2MASS position (obtained from NED, http://nedwww.ipac.caltech.edu/) to locate the nucleus. We find no obvious nuclear source in NGC 3377 and NGC 3923. The nearest source is 1.4” (1.98”) off from the 2MASS position of NGC 3377 (NGC 3923), which is considerably larger than the error (< 0.5”) of the on-axis Chandra source centroid (e.g., Kim et al. 2006). To extract the source spectra, we use a circle with a radius of 2.5” corresponding to 95% Enclosed Energy (EE) or better at E < 3 keV. If necessary, we increase the radius for a bright nuclear source. If there are nearby sources overlapping with the nuclear source region, we manually adjust the overlapping regions to properly exclude their emission. Properly choosing the region to extract the nuclear emission is important not only to measure the nuclear properties, but also to exclude the nuclear emission for accurate measurement of hot gas properties.

Since the X-ray emission from the hot gas is peaked toward the center (sometimes more steeply than the optical light), the hot gas may contribute significantly to the X-ray emission of the nuclear region, particularly for those galaxies with weak nuclei. The contribution from a population of ABs + CVs is generally small, but still non-negligible in gas-poor galaxies with a weak nucleus. Therefore, we fit the nuclear spectra with a combination of two PL + two APEC models. One PL represents the nuclear emission and one APEC the gas emission. The 2nd set of APEC + PL represents a population of ABs + CVs with their normalizations fixed at the corresponding $L_K$ which is determined within the source region. Although the gas temperature from these fits is not well constrained in most galaxies, it is generally close to that determined from the fit of the spectra from the diffuse emission (Section 3.4). Therefore, we set the gas temperature to be the same as found in the diffuse emission regions. The fitting results are listed in the 1st row of Table 3. There we show the $\chi^2$ and degrees of freedom in the fit, the temperature T of the hot gas determined from fitting in the diffuse region (DIFF), the power law slope gamma determined from fits to the AGN region, the $L_X$ of the power law component or 7 keV Bremsstrahlung component determined by fits in the 3 regions, the $L_X$ of the gas component, and the $L_X$ of the APEC and PL components from stellar emission scaled by K magnitude of the region (fixed for each fit). The fit is generally good with reduced $\chi^2$ close to 1. The best fit PL slope ranges from 1-2.2 which is typical for AGNs. Two exceptions are two strongest nuclei in NGC 1052 and NGC 4261. The best fit PL indices are negative in both cases, because they require more complex emission models than a single PL for the nuclear emission and extra absorption (e.g., Gonzalez-Martin et al. 2009). However, our measured luminosities of these two nuclei are still consistent with those in Gonzalez-Martin et al. (2009). In some cases we could not fit the AGN spectrum because of a small number of counts. In these cases we fixed the power law index to 1.8 and subtracted an estimate of the gas luminosity by scaling the count rate in an annulus of the diffuse region surrounding the AGN region. We consider the AGN luminosities measured in this way to be upper limits.

We expect a contribution from unresolved LMXBs to the emission in the central region, which cannot be modeled separately, because its hard X-ray spectrum is similar to that of the nucleus. LMXBs can be fit with either a power law or thermal Bremsstrahlung (see Section 3.3). Based on the $L_K$ ratio between the central region and the entire galaxy, we expect that up to 3 - 6% of $L_X$ (LMXB) could be unresolved in the central region, contaminating our estimates of the X-ray luminosities of very weak nuclei (e.g., NGC 3379, NGC 4697). In that case, $L_X$ (nucleus) should be considered as an upper limit.

## 3.3. Detected Low mass X-ray binaries (LMXBs)

Using *wavdetect* source positions, we extract the X-ray spectra of detected LMXBs from circular regions with a radius of 2.5” or 95% EE at 1.5 keV, whichever larger. The X-ray spectra of LMXBs have been studied previously (e.g., Irwin et al. 2003, Kim & Fabbiano 2003). More detailed studies of individual sources, including flux and spectral variations can be found in Fabbiano et al. (2010) and Brassington et al. (2010). Since our primary concern is to measure the total X-ray luminosity of LMXBs, we only look for the best parameter to represent the entire population of LMXBs.

We first fit the LMXB spectra with PL or thermal Bremsstrahlung (BR) models, to establish a template for this emission. The resulting best-fit parameters are $\Gamma$ = 1.4-1.8 for PL and kT = 5-10 keV for BR. For both models, the goodness of the fit is reasonable with a reduced $\chi^2$ close to 1. We note that the BR model fits slightly better (10-20% lower in total $\chi^2$) than PL, particularly for galaxies with the best statistics (largest counts). In either case, the resulting luminosities are identical in the soft energy band (0.3-2 keV). However, in the hard energy band (2-8 keV), BR produces systematically lower $L_X$ than PL, because of the steeper exponential decline toward higher energies in BR. In the broad 0.3-8 keV band, $L_X$ (BR) is lower by 10%. Given its better statistics, we take the BR model with kT fixed at 7 keV to represent the spectrum of LMXBs. We note that our results do not change within the uncertainties, if we adopt the PL model.

To determine the contribution of the other emission components to the $L_X$ of the detected LMXB regions, we apply a combination of four emission components: to the BR of the LMXB emission (with kT=7 keV) we add an APEC component for modeling the gas emission, and the set of APEC + PL best representing the ABs + CVs spectrum (see Section 3.1 and Appendix A) with their normalizations fixed at the $L_K$ determined within the LMXB region. As in Section 3.2, the gas temperature, while not well constrained in most galaxies, is close to that determined from the diffuse emission (Section 3.4). We set the gas temperature to be the same as that in the diffuse emission. The fitting results are listed in the 2$^{nd}$ row of Table 3 for each galaxy. The fit is good in all galaxies with reduced $\chi^2$ close to 1.

The ABs + CVs contribution in the LMXB region is considerably lower than that of LMXBs, since $L_X$(AB+CV) from the entire galaxy is ~10 times lower than Lx(LMXB) (see section 4). The contribution from the hot gas varies widely in different galaxies. $L_X$(gas) from the LMXB region is typically less than 10% of $L_X$(gas) from the diffuse emission region, but it can be higher for gas-poor galaxies when a large fraction of the central region is included in the LMXB region (e.g., NGC 1023 and NGC 3379).

## 3.4. Diffuse Emission

The diffuse emission is extracted from a circular region centered on the galaxy center from which all detected point sources (as described in Sections 3.2 and 3.3) are excluded. The outer radius is the point where the diffuse emission reaches the background level determined by examining the radial profile of the diffuse emission, and varies from galaxy to galaxy. Because bright LMXBs ($L_X > 10^{38}$ erg s$^{-1}$) in our sample galaxies are mostly detected, the contribution from unresolved LMXBs to the total $L_X$ is relatively small. However, the exact amount of unresolved LMXBs is still important for measuring the luminosity, temperature, and metal abundances of the hot ISM. To establish this contribution, we followed the two different approaches described below, which give consistent results.

### 3.4.1 Multi-component spectral fitting

Since the diffuse emission consists of hot gas, unresolved LMXBs and ABs+CVs, we model the spectra with a combination of four emission components: APEC for gas, BR for LMXBs and a set of APEC + PL for ABs + CVs. The temperature of BR is fixed at 7 keV (see Section 3.3). The normalizations of APEC + PL are again fixed for the corresponding $L_K$ determined in the region of the diffuse emission (Section 3.1). The temperature kT and power law slope $\Gamma$ for the APEC and PL components are also fixed as given in Section 3.1 (and Appendix A).

The fitting results are listed in the 3rd row of Table 3. The fit is good in most galaxies with reduced $\chi^2$ close to 1, except for NGC 4472 and NGC 4649 (see below). The temperature of the hot gas is usually well determined with a relatively small error even in the gas-poor galaxies. It ranges from 0.2 - 0. 8 keV and an error is typically 10-20%. However, the metal abundance is not well constrained, in most gas poor elliptical galaxies. We fix the abundance at the solar abundance (except for NGC 4472 and NGC 4649). We also test with variable abundances, but that does not significantly change $L_X$ (gas).

The diffuse spectra from NGC 4472 and NGC 4649, the two galaxies with the largest amount of the hot ISM in our sample, are not well reproduced (reduced $\chi^2$ ~ 3-5 for 250-270 dof) by the above simple model which assumes the gas is isothermal and all metal elements are solar. We allow individual elements to vary independently. The gas temperature also varies in different regions (increases with increasing distance from the center in both galaxies). Here, we present the (flux-weighted) average temperature and luminosity of the hot ISM and will discuss the detail gas structures and abundance measurements of different elements in a separate paper.

In most galaxies, undetected LMXBs contribute only a small fraction (< 25%) of the total luminosity of LMXBs, consequently the error in the luminosity of undetected LMXBs does not affect much the total luminosity of LMXBs. The fraction of undetected LMXBs is higher than 25% only in four galaxies. The two gas-rich galaxies NGC 4472 and NGC 4649 have fractions of 30-50% because the large amount of extended diffuse emission makes it hard to detect faint LMXBs. The two galaxies with the strongest nuclei, NGC 1052 and NGC 4261, have 40-50%, based on the spectral fitting. However, this is partly because of the emission from the PSF wing of the nuclei. Since the luminosity ratios of undetected LMXBs to nuclei are about 4%-15%, a small fraction of nuclear emission could significantly affect the luminosity of undetected LMXBs when measured from the diffuse emission.

Accurate measurements of the contributions of both undetected LMXBs and ABs+CVs are important in our sample, because these luminosity are not negligible compared to $L_X$(gas). In 14 galaxies, the luminosity ratio of undetected LMXBs to hot gas in the diffuse emission is > 25%, and in seven of them, the X-ray luminosity of undetected LMXBs is comparable to or greater than that of the hot ISM. In nine galaxies, we find $L_X$(gas) < $L_X$(AB+CV). It is important to note that both gas temperature and luminosity in galaxies with a small amount of hot gas [$L_X$(gas) < $10^{39}$ erg $s^{-1}$ and kT < 0.4 keV] would have been found spuriously higher, ignoring the contribution of undetected stellar sources.

### 3.4.2 Extrapolating the LMXB XLF

The XLF is relatively well known down to $L_X=10^{37}$ erg $s^{-1}$ for a few galaxies with ultra-deep Chandra observations (e.g., Kim et al. 2009; Voss et al. 2009). While the XLF in the entire $L_X$ range may be characterized by multiple power-laws (see Figure 3 in Kim & Fabbiano 2010), one of the key feature is that the XLF shape is more or less fixed with a universal slope of ~1 (in the form of d N / d ln $L_X$) between $L_X$ = 5 x $10^{37}$ erg $s^{-1}$ and 5 x $10^{38}$ erg $s^{-1}$ (Kim & Fabbiano 2004; Gilfanov 2004). We can utilize this feature to extrapolate $L_X$ from unresolved faint LMXBs, based on completely detected bright LMXBs.

First, we determine the ratio of the X-ray luminosity between bright and faint LMXBs using the ultra deep Chandra observations of NGC 3379, NGC 4278 (Brassington et al. 2008, 2009) and NGC 4697 (Sivakoff et al. 2007), which were observed with Chandra for 320, 460, and 130ks, respectively. The source detection limit (at a confidence level of 90%) are 6 x $10^{36}$ for NGC 3379 and 1.4 x $10^{37}$ erg $s^{-1}$ for NGC 4278 and NGC 4697 (Kim et al. 2009). Then we apply this ratio to estimate the contribution from the undetected LMXBs in other galaxies. For this purpose, we define a luminosity ratio $R_{15}$ = $L_X$(LMXBs with Lx < 5 x $10^{38}$ erg $s^{-1}$)/ $L_X$(LMXBs with $L_X$ = 1 - 5 x $10^{38}$ erg $s^{-1}$). The lower $L_X$ limit in the denominator corresponds to the detection limit at the Virgo cluster distance for a Chandra exposure of 40-50 ksec. We do not use very luminous LMXBs with Lx > 5 x $10^{38}$ erg $s^{-1}$, where the XLF becomes considerably steeper (KF04; Gilfanov 2004). Because of this XLF break, the very luminous LMXBs are relatively rare and a small number of luminous sources can significantly affect the ratio. Moreover, the relative fraction of very luminous LMXBs varies, depending on the stellar age of the parent galaxy (Kim & Fabbiano 2010). At lower luminosities ($L_X$ < 5 x $10^{37}$ erg $s^{-1}$), LMXBs in the field and in globular clusters (GC) have different XLF slopes (flatter in GC LMXBs: Kim et al. 2009, Voss et al. 2009), implying that the XLF may vary depending on different proportions of field and GC LMXBs. However, the contribution of these fainter LMXBs to the integrated $L_X$(LMXB) is small, and $R_{15}$ is not affected substantially.

That the exclusion of the fainter binaries does not affect significantly our results is demonstrated by the local dwarf elliptical galaxy M32, where at the distance of 0.8 Mpc LMXBs can be completely detected down to $L_X$ = 9 x $10^{33}$ erg $s^{-1}$. Of 22 sources detected inside the $D_{25}$ ellipse, only two sources are more luminous than $L_X$ = $10^{37}$ erg $s^{-1}$. However, the total $L_X$ of the 20 faint LMXBs is only 4% of the total $L_X$(LMXBs). The total $L_X$ of faint LMXBs with $L_X$ < 1×$10^{37}$ erg $s^{-1}$ in M32 is $L_X$(LMXB < 1×$10^{37}$ erg $s^{-1}$) = 3.8 x $10^{36}$ erg $s^{-1}$. If we scale it to that appropriate for NGC 3379 (using the ratio of the K-band luminosity of this galaxy and M32, see KF04), we obtain $L_X$(LMXB < 1×$10^{37}$ erg $s^{-1}$) = 1.9 x $10^{38}$ erg $s^{-1}$. In NGC 3379, the detected faint LMXBs with $L_X$ < 1 x $10^{37}$ erg $s^{-1}$ already contribute to $L_X$ = 1.82 x $10^{38}$ erg $s^{-1}$, suggesting that the remaining LMXBs could contribute only to $L_X$ = 8 x $10^{36}$ erg $s^{-1}$. In this case, $L_X$ of faint LMXBs with $L_X$ < 1 x $10^{37}$ erg $s^{-1}$ would be 3% of the total LMXBs or 5% of LMXBs with $L_X$ < 5 x $10^{38}$ erg $s^{-1}$. Given that NGC 3379 has the lowest LMXB detection limit ($L_X$ = 6 x $10^{36}$ erg $s^{-1}$ at 90%) among early type galaxies observed with Chandra, the contribution from the undetected LMXBs is quite small.

In the other two galaxies (NGC 4278 and NGC 4697) where the 90% detection limit is a factor of two higher ($L_X$ = 1.4 x $10^{37}$ erg $s^{-1}$), the X-ray emission from the undetected LMXBs would be slightly higher than that of NGC 3379, but the relative contribution from the undetected LMXBs to the total $L_X$(LMXB) remains small, because of the higher total luminosities of all detected LMXBs in these two galaxies (see Table 2). We note that the ratio of $L_X$(LMXB)/$L_K$ varies from one galaxy to another in our sample (see also Kim et al. 2009). This variation can be as much as a factor of 2 and depends most significantly on the GC specific frequency, $S_N$ (KF04; see Section 4.1). However, most faint LMXBs (with $L_X$ < a few x $10^{37}$ erg $s^{-1}$) are expected to be field LMXBs, because of the significant lack of faint GC-LMXBs (Kim et al 2009; Voss et al. 2009). Therefore, $L_X$ from faint LMXBs can be assumed to be fairly closely related to the K-band luminosity.

In Table 4, we list the number and total $L_X$ of LMXBs in different $L_X$ bins and $R_{15}$ for each galaxy; $R_{15}$ is almost identical in the three galaxies, ~ 1.9 and 2.0. This similarity justifies the applicability of this XLF method to other galaxies as long as luminous LMXBs with $L_X$ > 1 x $10^{38}$ erg $s^{-1}$ are detected. Combining LMXBs from all three galaxies, we obtain $R_{15} = 1.95 \pm 0.04$.

In our sample there are a few galaxies with detection limit slightly higher than $L_X$ = 1 x $10^{38}$ erg $s^{-1}$ (see Table 2). For this reason we also define a $R_{25}$ ratio by setting the lower $L_X$ limit at 2 x $10^{38}$ erg $s^{-1}$: $R_{25}$ can be applied to more galaxies, but it is subject to a larger error than $R_{15}$. We also list $R_{25}$ in Table 4. $R_{25}$ is similar in NGC 4278 and NGC 4697 and about twice of $R_{15}$, while its value is slightly lower in NGC 3397. Again, combining all LMXBs from three galaxies, we obtain $R_{25} = 3.80 \pm 0.97$.

We apply $R_{15}$ or $R_{25}$, as appropriate, to determine the X-ray luminosity of undetected LMXBs. In Figure 1, we compare the results from the spectral fitting and by extrapolating the XLF. In most galaxies, the two measurements agree with each other within the errors. The rms deviation from the equality (the diagonal line in Figure 1) is about a factor of 2.

## 3.5 Summary of X-ray luminosities from individual components

In Table 5, we summarize the X-ray luminosities from the different components (nucleus, AB+CV, LMXBs and hot gas) estimated from the results from the region (Table 3). We plot the X-ray luminosities against the K-band luminosity in Figure 2 where different components are marked by different symbols. The $L_X$-$L_K$ diagram of the total luminosity (marked by an open black circle), is similar to previous results (e.g., Eskridge et al. 1995; O'Sullivan et al. 2001; see also review in Fabbiano 1989). Now, in addition, we display the individual emission components in this $L_X$-$L_K$ diagram.

Since we estimate the contribution from ABs and CVs using a fixed $L_X(AB+CV)/L_K$ ratio (= 9.5 x $10^{27}$ erg $s^{-1}$ $L_{K\odot}^{-1}$ (Appendix A), $L_X$(AB+CV) is marked by a linear diagonal line in Figure 2. The X-ray luminosity of LMXBs (blue squares) is also proportional to $L_K$, but with a non-negligible scatter (see Section 4.1). The LMXB integrated luminosity is about 10 times larger than that of ABs + CVs: $L_X(LMXB) \simeq 10 \times L_X(AB+CV)$.

Our sample covers fairly uniformly the range of $L_X(gas) = 10^{38} - 10^{41}$ erg $s^{-1}$ and $L_X(gas)/L_K = 10^{27} - 10^{30}$ erg $s^{-1}$ $L_{K\odot}$, and is equally split among galaxies with $L_X(gas)<L_X(AB+CV)$; $L_X(AB+CV)<L_X(gas)< L_X(LMXB)$; and $L_X(gas) > L_X(LMXB)$. The X-ray luminosity of the hot ISM (red circles) is correlated with $L_K$ but with a larger scatter than the $L_X$(total) used as proxy for $L_X$(gas) in previous work (e.g., Eskridge et al. 1995; O'Sullivan et al. 2001; see Section 4.2).

The nuclear emission (green triangles) spans more than 2 orders of magnitude and does not seem to relate with $L_K$. We refer to Pellegrini (2010) for detailed discussions on the nuclear emission. Our results are generally similar to those of Pellegrini (2010), as we show in Figure 3. NGC 4649 is an outlier; our $L_X$(AGN) is significantly higher. The source of the measurement presented by Pellegrini (2010) is Soldatenkov et al. (2003), who detected the AGN only below 0.6 keV and extrapolated assuming $\Gamma$>2.2. Our measurement includes Chandra observations of NGC 4649 subsequent to Soldakenkov et al., which double the total exposure time. Our Lx(AGN) for NGC 4365 is also significantly higher than the value presented by Pellegrini, which is based on Gallo et al. 2010, who scaled the count rate to an X-ray luminosity assuming $\Gamma$=2. For both of these galaxies, we find a harder power law. The X-ray emitting gas in the AGN region is only a small fraction of the total emission and these uncertainties are not likely to affect our results. In the following Section, we will discuss how our results affect the understanding of the X-ray properties of LMXBs and hot gas.

# 4. Discussion

## 4.1 Low mass X-ray binaries

The linear relation between the integrated X-ray luminosity of the LMXB population and the stellar K-band luminosity of early-type galaxies is well established (e.g., White et al. 2002; Colbert et al. 2004; Kim & Fabbiano 2004; David et al. 2006). Given our full modeling of the X-ray emission components in a larger sample of early-type galaxies, we now revisit this relation.

The comparison of our results with those of KF04 is shown in Figure 4. With our new results, we find that the mean of log $L_X(LMXB)/L_K$ (in erg $s^{-1}$ $L_{K\odot}^{-1}$) = 29.0 $\pm$ 0.176; the standard deviation (1$\sigma$ rms) is 50%. The two horizontal cyan lines in Fig. 4a indicate the KF04 estimate of the 1$\sigma$ $L_X(LMXB)/L_K$ range and the two magenta lines indicate our new estimate. While the allowed ranges overlap, the average $L_X(LMXB)/L_K$ is now lower. We can understand this difference, by considering the characteristics of the two samples. First, KF04 selected 14 early type galaxies with a large number of detected LMXBs; this sample was selected to optimize the number of LMXBs and biased toward galaxies with a high $S_N$, since these GC-rich galaxies tend to have a larger number of LMXBs than GC-poor galaxies of the same $L_K$, and therefore their average $L_X(LMXB)/L_K$ is larger (White et al. 2002; KF04). Our new sample instead includes a reasonable coverage of optical galaxy properties; this sample includes a number of GC-poor galaxies and covers more uniformly the range of $S_N$, which have lower $L_X(LMXB)/L_K$ (see below). Second, the new sample excludes cD type galaxies, which tend to host a large number of GCs, for example, NGC 1399, which has the largest $L_X/L_K$ in KF04. In addition, KF04 estimated $L_X$(LMXB) by extrapolating down to $L_X=10^{37}$ erg $s^{-1}$ the XLF determined at $L_X > 5 \times 10^{37}$ erg $s^{-1}$, using a power-law model with slope of 2. We now know that the single power-law slope is flatter (~1.6) when determined with considerably deeper observations in the range $L_X = 10^{37}$ - $5\times10^{38}$ erg $s^{-1}$; here we have used this slope (Kim et al 2009; Voss et al. 2009; see Section 3.4.2).

Figure 4b shows the $L_X(LMXB)/L_K$ - $S_N$ relation from our sample. Again this relation is slightly shifted downward (to lower $L_x(LMXB)/L_K$) from that in KF04 for the same reasons described in the above. We find the best fit relation (solid line in Figure 4b): $L_X(LMXB)/\ L_K = 10^{28.88}$ x ${S_N}^{0.334}$ erg $s^{-1}$ $L_{K\odot}^{-1}$. The resulting p-value (or null hypothesis probability) is 0.005, indicating a strong correlation, applying a linear model fit available in the R package. The exponent of $0.334 \pm 0.106$ indicates that the difference in $S_N$ (between 0 and 8) could account for a factor of 2 spread in $L_X/L_K$. The remaining residual from the best fit is reduced to 40% in $1\sigma$ rms. This non-negligible residual may be partly because of the potential error in $S_N$, particularly in measurements with ground data.

## 4.2 Hot gas

### 4.2.1 The $L_X$(gas) – $L_K$ relation

A long standing puzzle in the X-ray study of early type galaxies is the two orders of magnitude spread in $L_X$(total) for a given optical luminosity; $L_X$ was used as a proxy for the hot gas content of the galaxies (e.g., Fabbiano 1989; White & Sarazin 1991; Eskridge et al. 1995; O'Sullivan et al. 2001; Ellis & O'Sullivan 2006; originally $L_B$ was used, now $L_K$ is preferred as a better proxy of the stellar luminosity). Several mechanisms have been proposed to account for this spread, including internal (e.g., dark matter, AGN feedback) and external effects (e.g., external confinement, ram pressure stripping, infall), but the proper physical process is yet to be explained (e.g., Fabbiano 1989; White & Sarazin 1991). The large $L_X$(total) / $L_B$ scatter was partly attributed to giant cD-type galaxies filling the high $L_X$ space in the $L_X$-$L_B$ plane (O'Sullivan et al. 2001). Since the hot gas dominates the X-ray emission in the latter, with $L_X$(gas) = $10^{41} - 10^{42}$ erg $s^{-1}$, $L_X$(gas) ~ $L_X$(total) in these galaxies. As we have shown above, in gas-poor early type galaxies, $L_X$(total) may still be $10^{40} - 10^{41}$ erg $s^{-1}$ because of the stellar contribution, but $L_X$(gas) is considerably lower: $L_X$(gas) = $10^{38} - 10^{39}$ erg $s^{-1}$ (see Figure 2). Therefore the true spread in the $L_X$(gas) – $L_K$ relation is larger than that of the $L_X$(total) – $L_K$.

We plot the $L_X$(gas) – $L_K$ diagram in Figure 5a. The average relations $L_X(LMXB)$ / $L_K = 10^{29}$ erg $s^{-1}$ $L_{K\odot}^{-1}$ and $L_X(AB+CV)$ / $L_K = 9.5 \times 10^{27}$ erg $s^{-1}$ $L_{K\odot}^{-1}$, are marked by two diagonal lines in this figure, dividing the diagram in three regions. Galaxies in these three regions are roughly divided by their gas temperature, kT > 0.4 keV, kT = 0.3–0.4 keV, kT < 0.3 keV, in the sense that the more luminous gaseous haloes are also hotter. Our sample covers a large range in $L_X$(gas) and $L_X$(gas)/$L_K$ including both gas-rich, intermediate, and gas-poor galaxies. This figure illustrates the importance of establishing the amount of stellar emission to determine accurately the gas properties of gas-poor galaxies.

As seen in Figure 5a, the spread is already more than 2 orders of magnitude in $L_X$(gas) for a given $L_K$ (~ $10^{11}$ $L_{K\odot}$). If we had included gas-rich cD galaxies in our sample, the spread in the $L_X$(gas) – $L_K$ relation would be even larger up to ~3 orders of magnitude. This brings an even bigger challenge for a proper theoretical explanation. Eskridge et al. (1995a) found a best fit slope between $L_X$(total) and $L_B$ of $1.8 \pm 0.1$ using the Einstein sample of early type galaxies. Similarly, O'Sullivan et al. (2001) found a best-fit slope of 2.2, using the ROSAT sample. In our sample, the linear relation between $L_X$(total) and $L_K$ is flatter (with a best fit slope of $1.4 \pm 0.2$) than the previous results, because gas-rich cD type galaxies are excluded by choice. However, it is clearly seen that the $L_X$(gas) - $L_K$ relation is steeper (best fit slope of $2.6 \pm 0.4$) than that with $L_X$(total) after the stellar contributions (from LMXBs, ABs and CVs) are removed.

In Table 6 we show the partial rank correlation coefficients and non-correlation probabilities among the 4 quantities $\sigma_*$, kT, $L_x$, and $L_K$. The partial rank coefficients (Kutner, Nachtsheim, & Neter, & Li 2004) test the significance of the correlation of two quantities while correcting for correlations with the others. Calculating the correlations of the ranks (the Spearman rank-order correlation coefficient) instead of the quantities themselves reduces dependence on the distributions of the quantities measured. The significance may be assessed by performing the 2-sided Student's t statistic test. The correlations between Lx(gas) and $L_K$ are very significant, both in the sample as a whole and restricted to brighter galaxies ($L_x > 10^{39}$ erg $s^{-1}$).

### 4.2.2 The $L_X$(gas) – $\sigma_*$ relation

The central velocity dispersion of early-type galaxies gives a measure of the central gravitational potential, but has also been related to and used as a proxy of the total galaxy potential. Correlations between $L_X$(total) of elliptical galaxies and $\sigma_*$ can be found in several papers in the literature (Eskridge et al. 1995a,b,c; Pellegrini, Held, & Ciotti 1997). Mahdavi & Geller (2001) found Lx(gas) ~ $\sigma^{10.2\ (+4.1,\ -1.6)}$, while Diehl & Statler (2005) found relations most consistent with Lx(gas) ~ $\sigma^{8.5}$. These correlations have been interpreted in terms of gravitational confinement of the hot ISM in the large gravitational potential of X-ray luminous ellipticals; outflows and winds were suggested to explain the X-ray faint ellipticals, which typically have lower $\sigma_*$ (e.g., Ciotti et al 1991).

We plot $L_X$(gas) against $\sigma_*$ in Figure 5b. $L_X$(gas) is well correlated with $\sigma_*$, although not as strong as with its relation with $L_K$ (see Table 6). Two most significant outliers are NGC 3115 and NGC 4621 (two galaxies in the lower-right corner in Figure 5b). They have very low $L_X$(gas) = 3 – 7 x $10^{38}$ erg $s^{-1}$ for their relatively high $\sigma_*$ ~ 260 km $s^{-1}$. The colors and morphology of NGC 3115 suggest its disk was a spiral that was swallowed by a much larger object (Michard 2007). NGC 4621 contains a counter-rotating core (Wernli, Emsellem, & Copin 2002), which could be related to mergers in the galaxy's history. Other galaxies with similar $\sigma_*$ typically have $L_X$(gas) ~ $10^{41}$ erg $s^{-1}$. Nonetheless, the correlation is best manifested by the lack of galaxies in the upper left corner, i.e., no galaxy with low $\sigma_*$ but high $L_X$(gas). In other words, all galaxies with a shallow potential depth (or $\sigma_*$ < 200 km $s^{-1}$) have only a small amount of the hot ISM ($L_X < 10^{40}$ erg $s^{-1}$). It is likely that they could not retain most of their hot ISM as the gas is in outflow/wind state (e.g., see Ciotti et al. 1991; Pellegrini & Ciotti 1998).

### 4.2.3 The kT-$L_K$ and kT-$\sigma_*$ relations

In Figure 6a and 6b, we plot the gas temperature against $L_K$ and $\sigma_*$, respectively. In general, this figure suggests a positive correlation between $L_K$ and $\sigma_*$; this is confirmed by the results of our statistical analyses. The partial rank correlation analysis (Table 6) confirms these correlations, with the weakest correlation between temperature and $\sigma_*$, particularly for gas-poor galaxies. The dashed line in Figure 6b indicates the relation that would arise if the gas temperature were fully determined by the stellar velocity dispersion: $kT_{gas} = kT_*$, where $kT_*=\mu\ m_H\ \sigma_*^2$. This line matches with the lower boundary in the $kT_{gas}$ - $\sigma_*$ plane, indicating that the gas energy is at least that associated with the stellar velocity dispersion, i.e. the gas is in thermal equilibrium with the stars. One exception is NGC 4526 which falls below the line in ~3$\sigma$ confidence (considering only the error in kT). The galaxies hosting large amounts of hot gas ($L_X$(gas) > 5 x $10^{39}$ erg $s^{-1}$, marked by red squares in Figure 6b) follow a similar slope of $kT_{gas}$ = $kT_*$, but they are shifted above the line by a factor of 1.5-2, indicating that they obtained additional energy input, roughly proportional to $\sigma_*$ (and likely $L_K$). This additional heating could be provided by SNe and AGN (Canizares et al 1987). Instead, in galaxies with a relatively small amount of hot gas ($L_X$(gas) < 5 x $10^{39}$ erg $s^{-1}$) we do not find a kT-$\sigma_*$ correlation (Table 6). For example, the eight galaxies (open squares in Figure 6b) with $L_X$(gas) = 1 – 5 x $10^{39}$ erg $s^{-1}$ have gas with almost identical temperature (0.3-0.4 keV), while $\sigma_*$ ranges from 160 to 250 km $s^{-1}$. The same is true for galaxies with the lowest $L_X$(gas) (< $10^{39}$ erg $s^{-1}$, marked by blue squares in Figure 6b), although the uncertainties in $kT_{gas}$ are large in this group. This lack of kT-$\sigma_*$ correlation in galaxies with small amounts of hot gas is consistent with their ISM being in a different physical state than in gas-rich galaxies. These galaxies, as previously suggested (e.g., Ciotti et al. 1991) may not be able to confine gravitationally their hot gas.

### 4.2.3 The $L_X$- kT relation

One of the most striking results is a positive correlation between the luminosity and temperature of the hot gas. As discussed above, the more luminous the hotter the gas is (see Figure 7). This relation is rather steep and the best fit relation is $L_X$(gas) ~ $T^{4..6 \pm 0.7}$ (green line in Figure 7). Since the gas parameters in extremely gas-poor galaxies with $L_X$(gas) ~ $10^{38}$ erg $s^{-1}$ are subject to a larger error, we also fit with only galaxies with $L_X$(gas) > $10^{39}$ erg $s^{-1}$. The exponent is similar (4.5 $\pm$ 0.55) in this selected sample (cyan line in Figure 7). In both cases, the null hypothesis probability is less than $10^{-6}$. The partial rank correlation analysis also confirms that the $L_X$– kT relation is one of the two strongest relations, the $2^{nd}$ one being $L_X$ — $L_K$ (in Table 6).

It is well known that the X-ray luminosity and the gas temperature are strongly correlated in bright clusters/groups of galaxies. For example, using HEAO-1 A2 data, Mushotzky (1984) showed $L_X \sim T^3$ among clusters of galaxies with $L_X = 5 \times 10^{43} - 3 \times 10^{45}$ erg $s^{-1}$ and kT = 2 – 9 keV. Using ROSAT observations of X-ray luminous early type galaxies (mostly brightest group/cluster galaxies), O'Sullivan et al. (2003) reported a similar relation. However, the relation between the gas luminosity and temperature has not been well established in the gas-poor early type galaxies, mainly because of limited understanding of gas properties in these systems (e.g., David et al. 2006).

While the slope of the $L_X$(gas) – T(gas) relation in our sample is consistent with 4.8 $\pm$ 0.7 measured by O'Sullivan et al. (2003), their best fit line (yellow line in Figure 7) is shifted up in $L_X$(gas) by an order of magnitude. This may be partly because of the difference in sample galaxies as $L_X$(gas) is higher in the cD type group/cluster dominant galaxies (majority of their sample) than non cD-type galaxies (our sample). However, the luminosity difference remains in T = 0.3- 0.6 keV or $L_X = 10^{39}$ - $10^{41}$ erg $s^{-1}$ where two samples overlap. We compare the $L_X$-T relation with 10 galaxies in common. Two most significant discrepant cases are NGC 4365 and NGC 4649. Both of them are obvious outliers from their mean relation. In NGC 4365, their kT = 1.0 (-0.2, +0.3) keV is too high for our T= 0.44 $\pm$ 0.02 keV. The higher temperature may be due to the incomplete subtraction of the hard emission from LMXBs in analyzing ROSAT data. This can be compared with the gas temperature of 0.56 keV (-0.08, +0.05) measured by Sivakoff et al. (2003) with early Chandra data. Note that Sivakoff et al. (2003) did not consider the contribution from ABs and CVs. In NGC 4649, their $L_X$ is lower by a factor 100 than ours. NGC 4649 is well known to have an extended hot ISM (e.g., Fabbiano, Kim & Trinchieri 1992). For the remaining galaxies (excluding NGC 4365 and NGC 4649), $L_X$(gas) is generally higher than ours (after correcting for different distances, as they adopted *Ho*=50 km $s^{-1}$ $Mpc^{-1}$), while the gas temperature is more or less consistent with our results. This may be partly because LMXBs are not properly excluded and partly because they did not consider the contribution from ABs and CVs.

David et al. (2006) also presented the $L_X$(gas) – T(gas) relation with Chandra data of 18 low luminosity early type galaxies, but could not find any clear correlation. Again we compare their results, using 9 galaxies in common. In contrary to the comparison with O'Sullivan et al. (2003), while the luminosity agrees well, the temperature is different (higher than our results) in a few galaxies. This may be partly because they did not consider the contribution from ABs and CVs. The most significant discrepancies are in NGC 1023 and NGC 3379. Although their errors are large in both cases, these two galaxies are the two most significant outliers (too high T for a given $L_X$) in their plot. We note that they used only the first observations in both galaxies and the data we use in this study are about 10 times deeper (see Table 2).

To illustrate these comparisons, in Figure 8 we show the galaxies in common with two previous studies: 10 galaxies (magenta open circles) common with O'Sullivan et al. (2003) and 9 galaxies (red open squares) with David et al. (2006). Our results are marked by blue filled circles. Two galaxies (NGC 4697 and NGC 4552) are also common in both samples. While $L_X$ and T are all consistent for NGC 4552, both $L_X$ and T from O'Sullivan et al. (2003) are quite different in NGC 4697 (see Figure 8). We separately mark those with significant discrepancies and link them with arrows. It is quite clear that once corrected, those apparent outliers in the previous studies do indeed nicely follow the general trend between $L_X$(gas) and T(gas).

Following our relation between the luminosity and temperature of the hot ISM ($L_X \sim T^{4.5}$), we find approximately that:

$L_X(gas) = 10^{38} – 10^{39}$ erg s$^{-1}$ for kT = 0.2 – 0.3 keV,

$L_X(gas) = 10^{39} – 10^{40}$ erg s$^{-1}$ for kT = 0.3 – 0.4 keV,

$L_X(gas) = 10^{40} – 10^{41}$ erg s$^{-1}$ for kT = 0.4 – 0.7 keV,

$L_X(gas) > 10^{41}$ erg s$^{-1}$ for kT > 0.7 keV,

Also in terms of $L_X(gas) / L_K$, we find approximately that (see the diagonal lines in Figure 5a):

$L_X(gas) / L_K < 10^{28}$ erg s$^{-1}$ $L_{K\odot}^{-1}$ for kT < 0.3 keV,

$L_X(gas) / L_K = 10^{28} – 10^{29}$ erg s$^{-1}$ $L_{K\odot}^{-1}$ for kT = 0.3 – 0.4 keV,

$L_X(gas) / L_K > 10^{29}$ erg s$^{-1}$ $L_{K\odot}^{-1}$ for kT > 0.4 keV,

Note that $L_X(ABCV)/L_K = 9.5 \times 10^{27}$ erg s$^{-1}$ $L_{K\odot}^{-1}$ and $L_X(LMXB)/L_K = 10^{29}$ erg s$^{-1}$ $L_{K\odot}^{-1}$.

To better understand the strong positive correlation between $L_X$ and T, we divide our sample into 3 groups by $\sigma_*$ and mark them differently in Figure 7: red squares for $\sigma_* > 240$ km s$^{-1}$, black open squares for $\sigma_* = 200$ -240 km s$^{-1}$ and blue squares for $\sigma_* < 200$ km s$^{-1}$. The positive $L_X$-T correlation holds in all three sub-groups as well as in the entire sample. The galaxies in the first group with the highest $\sigma_*$ would be able retain most of their ISM, compared to the other groups with lower $\sigma_*$. We can qualitatively understand this correlation because the larger galaxy retains a larger amount of the hot ISM and more energy (by mass loss from evolved stars and SNe) was added to the ISM. The correlation is likely a scaled-down version of similar relations found in cD galaxies (O'Sullivan et al. 2003) and groups and clusters of galaxies (e.g., Mushotzky 1984). However, the exact relation, $L_X \sim T^{4.5}$, needs to be explained.

In the middle group with intermediate $\sigma_*$ (200 – 240 km s$^{-1}$), the general positive correlation remains the same. However, they may form an S-shape in the $L_X$-T diagram. This is most clearly visible by a significant $L_X$ drop (by a factor of ~20) among seven galaxies with a narrow range of kT. These seven galaxies have similar kT (0.32-0.36 keV) and similar $\sigma_*$ (202-232 km s$^{-1}$), but significantly different $L_X$(gas). They are NGC 1023, 1052, 1549, 2768, 3585, 4278, and 4473. To double-check whether $L_X$ is really unrelated with $\sigma_*$ even though they are in a narrow range of $\sigma_*$, we check $L_X$ and $\sigma_*$ for these galaxies. They are also the intermediate group (kT = 0.3 — 0.4 keV) in the $L_X$-$\sigma_*$ diagram (Figure 5b). There is no trend among these galaxies within $\sigma_*$=200-240 km s$^{-1}$. If this is real, what makes the scatter in the hot gas amount among these seven galaxies, even if their gas is in a similar temperature under the similar gravitational potential depth? If this sudden $L_X$ drop indicates a transition of the gas state from inflow to outflow, then what triggers the transition? Their K-band luminosity is also in a relatively narrow range (7-18 x $10^{10}$ $L_{K\odot}$) and seems to be unrelated with $L_X$(gas). This excludes any potential difference caused by the SNe energy input. We check whether AGN may be responsible for the $L_X$ drop among these seven galaxies. An additional energy input by the AGN feedback to the hot ISM could trigger the outflow. In this case, we may expect an anti-correlation between $L_X$(AGN) and $L_X$(gas). However, there is no such trend. Furthermore, we would also expect a positive correlation between $L_X$(AGN) and $T_{gas}$. But there is no such trend either, because they all have very similar $T_{gas}$.

Finally, we check whether the $L_X$ drop is related to the recent star formation triggered by minor/major mergers. As opposed to the typical old stellar system where most stars formed very early in a relatively short time scale, a considerable number of early type galaxies exhibit a signature of recent star formation episodes (e.g., Trager et al. 2000; Schweizer 2003). We take the average stellar age measured by the optical line indices (see section 2 and Table 1). We plot $L_X$(gas) and age in Figure 9. It is interesting to note that $L_X$(gas) may indeed correlate with age (with the null hypothesis probability of 0.14), in a sense that younger galaxies tend to have a smaller amount of gas. The 2$^{nd}$ generation star formation could add enough energy to the hot ISM so that these galaxies would have emptied their ISM. A small amount of the hot ISM may have been accumulated since the last star formation episode (see also Fabbiano & Schweizer 1995 and Kim & Fabbiano, 2003). On the other hand, old galaxies would have experienced the wind during the early star formation period, but they would have a longer time to accumulate the ISM by mass loss from the evolved stars. Although our result is based on a small sample, it is very encouraging and deserves to be confirmed with a larger sample. We note that age-related X-ray signatures are also reported in luminous LMXBs (Kim & Fabbiano 2010) and in metal abundance ratios (Kim 2010; Kim et al. in prep.)

The group with the lowest $\sigma_*$ (blue squares in Figure 7) also exhibits a positive correlation between $L_X$(gas) and T. However, this positive correlation is not easy to understand. Given that they would have shallower potential depth than the other groups with higher $\sigma_*$, their ISM is likely in the outflow/wind state where the gas pressure overcomes the gravitational potential. In this case, among galaxies with similar $\sigma_*$, the hotter gas under higher pressure would be in a stronger wind state which results in lower $L_X$(gas), i.e., the gas temperature is expected to be anti-correlated with the gas luminosity. What we are seeing is clearly the opposite. Using only galaxies in this group (but excluding M32 and NGC 821), we refit the relation and find a similar slope (4.9 $\pm$ 1.3) as in the full sample. The correlation is moderately strong with a null hypothesis probability of 0.013. $L_X$(gas) does not seem to be related with any other quantities, like $L_X$(AGN), $L_K$ and age.

In our sample, the lowest measureable temperature and luminosity go down to kT ~ 0.2 keV with $L_X$(gas) ~ $10^{38}$ erg $s^{-1}$. The galaxy with the least amount of the hot gas is M32. Since M32 is almost devoid of gas with an upper limit of $L_X$(gas) < 8 x $10^{36}$ erg $s^{-1}$, its gas parameters are not well determined. NGC 821 has also very little gas, if any (as shown in Pellegrini et al. 2007b) with $L_X$ (gas) = 2 x $10^{37}$ erg $s^{-1}$ (or $L_X$ = 0 - $10^{38}$ erg $s^{-1}$ in 1$\sigma$). Its temperature is 0.15 keV but with a large error (0.1 – 1.0 keV). NGC 3377 has the next lowest gas luminosity, $L_X$(gas) = 1.1 x $10^{38}$ erg $s^{-1}$ (or $L_X$ = 0.45 – 2 x $10^{38}$ erg $s^{-1}$ in 1$\sigma$) with kT = 0.25 keV (0.2 - 0.3 keV). Since M32 and NGC 3377 are the lowest in $\sigma_*$ (72 and 107 km $s^{-1}$, respectively), they are not able to hold their hot ISM. However, NGC 821, an isolated elliptical galaxy, has $\sigma_*$=189 km $s^{-1}$. Other isolated galaxies with comparable $\sigma_*$ (~180 km $s^{-1}$) typically have $L_X$(gas) = $10^{39}$ – $10^{40}$ erg $s^{-1}$ (see Figure 5b). Pellegrini et al. (2007) showed by hydrodynamical simulations that stellar mass losses could be driven out of NGC 821 in a wind sustained by Type Ia SNe. If so, it is hard to explain why other galaxies with similar parameters ($\sigma_*$, age, environment) retain a significantly larger amount of the hot ISM. Since NGC 821 is an old (9 Gyr) elliptical galaxy, age does not seem to be an important factor. The nucleus of NGC 821 is inactive with $L_X \sim 10^{39}$ erg $s^{-1}$. There may be a jet (Pellegrini et al. 2007b), indicating some nuclear activities in the past, but it does not seem to be strong enough to distinguish NGC 821 from other galaxies.

Another possibility is that the stellar velocity dispersion is not a good indicator of the potential depth, because it could be affected by the galaxy rotation and/or anisotropic stellar orbits (e.g., Scott et al. 2009). However, the mass of the dark matter in the central region is only a fraction of the total mass. For example, among the SAURON sample, the median dark matter fraction is about 30% of the total mass inside one effective radius (Cappellari et al. 2006). Even if the dark matter faction (or mass to light ratio) varies from one galaxy to another, it is still proportional to the galaxy size and $\sigma$ as shown in the SAURON study (Cappellari et al. 2006; Scott et al. 2009) and the Sloan Lens ACS survey (Auger et al. 2010) such that $\sigma$ is still a good indicator of the total mass.

## 5. Summary and Conclusions

Selecting a sample of 30 early type galaxies with deep Chandra observations and optical spectroscopy, we measure the X-ray properties of individual sources (AGN, gas, and LMXB) and compare with other basic galaxy properties. In summary we find:

1. Our sample covers a wide range in $L_X$(gas) and $L_X$(gas)/$L_K$. In 1/3 of our sample, $L_X$(gas) is lower than $L_X$(ABs+CVs). The contribution from undetected stellar X-ray sources needs to be properly accounted, particularly to accurately measure gas properties in gas-poor galaxies.

2. Considering the contribution from the undetected LMXBs by fitting the spectra of the diffuse emission (after excluding all detected point source) and as well as by extrapolating X-ray luminosity function of LMXBs, we revise the relation between $L_X$(LMXB) and $L_K$:

$$L_X\ (LMXB)\ /\ L_K = 10^{29.0 \pm 0.176} \quad \text{erg s}^{-1}\ L_{K\odot}^{-1}$$

This is consistent with the previous results in KF04, but slightly lower because of our sample covering more uniformly in $S_N$ and inaccurate XLF extrapolation applied in KF04. Considering the dependence of the GC specific frequency ($S_N$), we find an improved relation:

$$L_X(LMXB)/\ L_K = 10^{28.88} \text{ x } {S_N}^{0.334} \quad \text{erg s}^{-1}\ L_{K\odot}^{-1}$$

3. On average, the X-ray luminosity of LMXBs is about ten times of that of ABs+CVs, i.e.,

$$L_X\,(LMXB) = 10 \text{ x } L_X(AB+CV).$$

4. Using $L_X$(gas) in place of $L_X$(total), we revise the $L_X$-$L_K$ diagram. We find that the wide range in $L_X/L_K$ Is even larger and that the best fit slope in the $L_X$-$L_K$ relation is steeper, because of adding more gas-poor galaxies for which $L_X$(gas) was not accurately measured. In particular, the long standing puzzle for the large span in $L_X$ among galaxies with similar $L_K$, $\sigma_*$, environment, and AGN remains unknown. Even larger spread in $L_X$(gas)/$L_K$ brings an even bigger challenge for a proper theoretical explanation

5. We find a positive correlation between the luminosity and temperature of the hot ISM with the best fit relation of $L_X \sim T^{4.5}$, when determined in the entire sample. This correlation also holds in three sub-groups binned by $\sigma_*$. Among galaxies with high velocity dispersions, this relation may be a continuation of similar relations found in more luminous cD-type galaxies and groups/clusters of galaxies.

6. We find an S-shape in the $L_X$-$L_K$ relation among galaxies with intermediate $\sigma_*$. Among galaxies with similar kT (0.32-0.36 keV) and similar $\sigma_*$ (202-232 km $s^{-1}$), $L_X$(gas) drops by a factor of ~20. This may be due to a transition of the gas state from inflow to outflow. Among these galaxies, we find no trend associated with $L_K$ and AGN. However, we find a weak, positive correlation between $L_X$(gas) and the average stellar age, possibly suggesting rejuvenated star formation may be responsible for this transition.

7. The positive $L_X$-T correlation is still moderately strong among galaxies with low velocity dispersions. Because the hot gas under the shallow potential depth in these galaxies is expected in an outflow/wind state, the $L_X$-T relation is expected to be negative (i.e., the hotter the gas, the stronger the wind is). This remains to be explained and points to the need for more theoretical work.

# APPENDIX A

## X-RAY EMISSION FROM ACTIVE BINARIES AND CATACLYSMIC VARIABLES

The X-ray emission from the Galactic stellar sources such as active binaries (AB, e.g., RS CVn) and cataclysmic variables (CV, initially called nova) has been known from the early X-ray missions (e.g., see Charles & Seward 1995). Their contribution to the X-ray luminosity of elliptical galaxies was estimated (e.g., Pellegrini & Fabbiano 1994), but often ignored because of their relatively weak luminosities, particularly when compared to more luminous LMXBs (see a review by Fabbiano 2006). With the high spatial resolution Chandra observations, most bright LMXBs are detected in nearby elliptical galaxies. After excluding those detected LMXBs, the stellar emission is not negligible any longer in the remaining unresolved emission, particularly in gas-poor elliptical galaxies. In this case, without a proper consideration of the stellar emission, the hot ISM properties, if determined with the diffuse X-ray emission, may be seriously misleading.

Recently Revnivtsev et al. (2007a) revisited this issue. After removing point sources with $L_X > 10^{34}$ erg $s^{-1}$, they showed that the 0.3—7 keV X-ray image and radial profile of M32 follow closely the IR K-band image and profile from a few to ~100 arcsec, indicating that the remaining diffuse emission is indeed dominated by ABs and CVs. They also estimated the scaling from K band magnitude to the X-ray luminosity of stellar sources (ABs + CVs), but with relatively large uncertainties. In the solar vicinity, RXTE and *ROSAT* X-ray observations have resolved point sources (ABs and CVs) in the $10^{30}$-$10^{34}$ erg $s^{-1}$ range (Sazonov et al. 2006). The Galactic ridge X-ray emission in the 3—20 keV range observed with XTE is found to trace the stellar near IR brightness distribution as observed with COBE/DIRBE (Revnivtsev et al. 2006). Revnivtsev et al. (2007b) used a deep *Chandra* observation of a region of the Galactic plane to resolve point sources with luminosities of $10^{30}$-$10^{32}$ erg $s^{-1}$. Another region towards the galactic center allowed the X-ray luminosity function to be constrained above $10^{30}$ erg $s^{-1}$. Furthermore, $84 \pm 12\%$ of the Galactic diffuse X-ray emission could be resolved into point sources by concentrating on the 6.5 – 7.1 keV range containing a blend of iron emission lines (Revnivtsev et al. 2009).

ABs fall in several categories. From observations with the *ROSAT* All Sky Survey, Makarov (2003) cataloged the 100 brightest X-ray stars within 50 parsecs of the Sun. The pre-main sequence stars, post-T Tauri stars, and very young main-sequence stars that contribute in the solar neighborhood will not contribute to early-type galaxies. The remaining stellar emission sources classified as ABs include RS CVn systems, named after their prototype, which are typically synchronously rotating binaries with an evolved component and at least one star of type F, G, or K. X-ray spectroscopy of such systems in quiescent and flaring states show general agreement with variable 2-temperature thin gas emission components (kT ~0.6-1.0, 2.0-2.5 keV), for example V711 Tau/HR 1099 (Osten et al. 2004) and II Peg (Covino et al. 2000). ABs may also include binary systems of the type BY Dra, semi-detached Algols, and $\beta$ Lyr systems. BY Dra stars, a category similar to the RS CVns, may be single or double rapidly rotating dwarfs with active chromospheres. Algols typically are 3-4 times dimmer than RS CVn systems with the same orbital period (Singh, Drake, & White 1996). The RS CVn systems are the brightest of the stellar X-ray emitters in the solar neighborhood (Makarov 2003).

CVs are accreting white dwarf systems, and may be classified as either magnetic or non-magnetic based on whether the accretion is directed by the magnetic field of the white dwarf or flows through an accretion disk. About 25% of CVs are magnetic, and 63% of those are Polars with larger magnetic field and synchronous rotation of the white dwarf, while 37% are the asynchronously rotating Intermediate Polars, which are brighter. In addition to the hard component, a soft blackbody component with kT ~ 30eV (Vrtilek et al. 1994), and partially ionized absorption from the source may complicate the spectrum (Baskill, Wheatley, & Osborne 2005).

We parameterize the X-ray spectra of a population of ABs and CVs and measure $L_X/L_K$ in various energy ranges to be easily applicable to other galaxies. We use two local group galaxies, M31 (NGC 224) and M32 (NGC 221) where LMXBs are completely detected and excluded. Because M32 does not retain any detectable amount of hot ISM, the diffuse emission is fully dominated by ABs and CVs. This is the only galaxy where we can really isolate the stellar emission. Although much brighter (than M32), M31 is known to contain some hot gas (Bogdan & Gilfanov 2008, Liu, et al. 2010) which mostly emits at energies below ~1.5 keV. However the X-ray emission above 2.5 keV is dominated by LMXBs and other stellar sources (Li, Wang, & Wakker 2009). We find the best constraints by jointly fitting two spectra with the M32 spectrum being more useful at lower energies and the M31 spectra being more useful at higher energies.

All Chandra data were taken from the Chandra archive (http://cxc.harvard.edu/cda/). We only use the ACIS-S (S3 chip) data. We list the basic observation log including observation id and combined exposure times in Table A1. Also listed are Galactic line of sight $N_H$ taken from the NRAO survey (Dickey & Lockman 1990), distances from Tonry et al. (2001), source extraction radii and K-band magnitudes. We extract the source spectra from the central 60" for both M31 and M32. In the outer region of M32, the X-ray emission is dominated by the background. In the outer region of M31, there is still significant source emission, but the X-ray radial profile starts to deviate from the K-band radial profile, indicating that the X-ray sources associated with the disk may contribute (see Li, Wang, & Wakker 2009). K band images were obtained from the 2MASS Large Galaxy Atlas (Jarrett et al. 2003).

Since the X-ray spectra of ABs and CVs are different, we attempt to parameterize their X-ray emission separately with two emission models: APEC (APEC, Smith et al. 2001) for the coronal emission including lines from metal elements of ABs and power-law (PL) for the featureless hard emission of CVs. However, it is likely that the APEC component includes some CV emission and the PL component includes some AB emission. Based on high-resolution X-ray spectroscopy of individual RS CVn systems, it is known that their spectra fit with two-temperature models including a harder (2.0-2.5 keV) thermal component (Covino et al. 2000, Osten et al. 2004) and the brighter AB tends to have a harder spectrum (Sazonov et al. 2006). Similarly, the power-law fit to the CV spectrum may be an oversimplification, particularly in the soft X-ray range (Vrtilek et al. 1994; Baskill, Wheatley, & Osborne 2005). Since our primary goal is to determine the total contribution from the stellar emission in entire galaxies, we collectively measure their X-ray emission from the total ABs + CVs, without distinguishing them. Therefore, we only consider that the combination of APEC and PL models represent the emission from the entire population of ABs and CVs.

We adopt GRSA solar abundances (Grevesse & Sauval 1998) for the APEC model. We allow the temperature and normalization to vary as free parameters. We also allow the abundance to vary, but keep the relative ratios to solar. For the PL model, we allow the photon index and normalization to vary as free parameters. $N_H$ is fixed at the line of sight Galactic value.

First, we fit M32 and M31 spectra individually. For M32, we fit the 0.3-5.0 keV spectrum with the absorbed APEC + PL model. In Figure A1, we show the observed spectrum with the best fit model. The APEC and PL components are also plotted separately. The PL component dominates at the high energies (> 2 keV), while the APEC component peaks at ~0.8 keV. In Table A2, we list the best fit parameters with corresponding errors and resulting statistics. The reduced $\chi^2$ is 0.8 for 136 degrees of freedom (dof), indicating a reasonably good fit (see also $\Delta\chi$ in the bottom panel of Figure A1). However, statistical errors (at $1\sigma$ confidence) are relatively large. While the photon index ($\Gamma$ ~ 1.8) and temperature (kT ~ 0.5 keV) are determined within 20-30%, the normalizations of these two components (expressed by $L_X/L_K$ in Table A2) are poorly constrained. We note that the abundance (often 10-20% solar) in the APEC model (in Table A2) is not an accurate measurement of the abundances in the stellar coronae because of systematic uncertainties in separating the AB and CV contributions to the spectrum. Audard et al. (2003) measured abundances from XMM-*Newton* spectra of 5 RS CVn systems and found from 3-temperature APEC fits and found values from 0.1 to 2.1 of solar abundance, with the mean for each system below solar.

Fitting the M31 bulge spectrum is more complex, because the hot gas significantly contributes at low energies (Bogdan & Gilfanov 2008, Li & Wang 2007; Li, Wang, & Wakker 2009) and because the temperature of the hot gas (0.3-0.4 keV, Li & Wang 2007) is similar to that of the stellar APEC component (~0.5 keV). Instead of fitting over the entire 0.3--5.0 keV range with a three component model (two stellar components + one gas component), we apply an iterative procedure. We first fit the hard X-rays (2.3 – 5.0 keV) to separately determine the PL component since both the soft stellar component and the gas component contribute less in this energy range, although not negligible ($\Gamma$ still depends on residual contributions from the thermal models). With fixed $\Gamma$ as determined from the fit in the 2.3 – 5.0 keV range, we fit the spectrum from 0.3—5 keV by adding two APEC models for the soft stellar and gas components. We then fix these models and re-fit the spectrum in the 2.3--5.0 keV range to re-determine the PL parameters. We repeat these iterations until they converge. First we tie all elements in the hot gas to vary together at the fixed solar ratio. Given the high S/N spectrum of the M31 bulge, the fit is not acceptable with the reduced $\chi^2$ of 1.16 for 136 degrees of freedom (Table A2). This corresponds to the probability to reject a null hypothesis of 9%. There are also significant local deviations in $\Delta\chi$, most significantly at ~0.5 keV. To improve the fit, we allow all the elements in the hot ISM to vary independently. The reduced $\chi^2$ significantly decreases to 0.83 for 126 degrees of freedom (Table A2), indicating a good fit with no clear local deviations in $\Delta\chi$ (at the bottom panel of Figure A2). Unlike M32, the hot ISM indeed dominates at low energies (< 1.5 keV). Because of this, the stellar APEC component is not well constrained (the best fit corresponds to zero normalization).

Neither M31 nor M32 alone is sufficient to characterize the stellar X-ray spectrum. The M32 spectrum results in large uncertainties in normalizations and the PL photon index, while the M31 spectrum is dominated by the hot gas at E < 2 keV, rendering the measurement of AP parameters uncertain. Taking advantage of the two spectra (M31 data having a high S/N and the PL component dominating in higher energies and M32 data being free from the hot ISM), we jointly fit them to better constrain the stellar (AB+CV) parameters. We assume that the total X-ray luminosity of ABs and CVs is proportional to the stellar K-band luminosity, $L_K$. The APEC and PL normalizations of the two galaxies are linked such that the $L_X$ ratio is the same as their $L_K$ ratio. The results of our fits again depend on how we fit the gas in M31 (i.e., how we tie the individual elements). Again the fit is significantly improved by allowing the abundances of individual elements to vary independently. Applying the same iteration procedure described above, we obtain a good fit with the reduced $\chi^2$ of 0.8 for 266 dof. The fitting results are summarized in Table A3 and the observed spectra and best fit models are shown in Figure A3a and A3b. Each model parameters are relatively well determined: $\Gamma$ and kT in 10-20%. The normalizations of the APEC and PL components are also well determined in 15-30%.

The best fit parameters are $\Gamma$=1.76 $\pm$ 0.37, kT = 0.48 (-0.05, +0.07) keV and Z = 0.18 (-0.07, + 0.19). The PL photon index can be compared with that expected from CVs. Magnetic CVs have been found to have $\Gamma$ = 1.22 $\pm$ 0.33, while nonmagnetic CVs have been found to have $\Gamma$ = 1.97 $\pm$ 0.20 (Heinke et al. 2008). Our result is consistent with a mixture of the two types of CVs. Similarly, the APEC component is representing the soft emission from the stellar coronal emission and possibly the soft blackbody component of CVs. Again, we note that the best fit value of Z does not reflect the abundance in the stellar coronae.

For easy application to other galaxies, we convert the normalizations of the APEC and PL components to the X-ray to K-band luminosity ratios in multiple energy ranges and list them with corresponding errors in Table A4. In the frequently used energy ranges (0.5-2 keV and 0.3-8 keV), the total stellar (ABs+CVs) X-ray luminosity for a given K-band luminosity are:

$$L_X/L_K = 4.4^{+1.5}_{-0.9} \times 10^{27} \text{ erg s}^{-1} L_{K\odot} \text{ in } 0.5-2 \text{ keV} \quad (1)$$

$$L_X/L_K = 9.5^{+2.1}_{-1.1} \times 10^{27} \text{ erg s}^{-1} L_{K\odot} \text{ in } 0.3-8 \text{ keV} \quad (2)$$

Since the total X-ray luminosity of LMXBs is also proportional to the total stellar K-band luminosity (e.g., Kim & Fabbiano 2004), we can directly compare $L_X$(LMXB)/$L_K$ and $L_X$(AB+CV)/$L_K$. On average, the population of ABs and CVs contribute about 1/10 of that from LMXBs (section 4).

In Table A4, we also compare our results with previous measurements. Our result is consistent with the previous result for M32 by Revnivtsev et al. (2007) and that for M31 by Bogdan & Gilfanov (2008), but lower than Revnivtsev et al. (2008) and Li & Wang (2007). In particular, Revnivtsev et al. (2008) value is higher by 60% in the soft energy range (0.5-2 keV) when determined with the Chandra data of NGC 3379, because they assumed no contribution from the hot ISM, although Trinchieri et al. (2008) identified the presence of hot ISM.

**Acknowledgments**

The data analysis was supported by the CXC CIAO software and CALDB. We have used the NASA NED and ADS facilities, and have extracted archival data from the *Chandra* archives. This work was supported by the *Chandra* GO grant G08-9133X (PI: Kim) and XMM-Newton GO grant NNX09AT20G (PI: Kim). This publication makes use of data products from the Two Micron All Sky Survey, which is a joint project of the University of Massachusetts and the Infrared Processing and Analysis Center/California Institute of Technology, funded by the National Aeronautics and Space Administration and the National Science Foundation. We thank Silvia Pellegrini for useful discussions.

**Figure Captions**

Figure 1. X-ray luminosities of undetected LMXBs measured by spectral fitting and by XLF are compared.

Figure 2. X-ray luminosities of individual components are plotted against the K-band luminosity. The total X-ray luminosity is denoted by open black circles, nuclei by filled green triangles, LMXBs by filled blue squares, hot gas by filled red circles.

Figure 3. Our measurements of Lx(AGN) compared with the values presented in Pellegrini (2010).

Figure 4. X-ray luminosity of LMXBs is plotted against (a) $L_K$ and (b) $S_N$. Two cyan lines indicate the 1$\sigma$ allowed range from Kim & Fabbiano (2004) and two magenta lines indicate the new result from this work. The blue linear line is the best fit, $L_X(LMXB)/\ L_K \ \sim\ {S_N}^{0.334}$.

Figure 5. X-ray luminosity of the hot gas is plotted against (a) $L_K$ and (b) $\sigma_*$. Three sub-groups in different kT bins are marked differently (red, black, blue in order of decreasing kT). The $L_X/L_K$ ratios corresponding to LMXBs and ABs+CVs are marked by two diagonal lines.

Figure 6. The gas temperature is plotted against (a) $L_K$ and (b) $\sigma_*$. Three sub-groups in different $L_X$(gas) bins are marked differently (red, black, blue in order of decreasing $L_X$). The cyan diagonal line indicates $kT_{gas} = kT_*$.

Figure 7. X-ray luminosity vs. temperature of the hot gas. Three sub-groups in different $\sigma_*$ bins are marked differently (red, black, blue in order of decreasing $\sigma_*$). Also over-plotted are the best fit relations determined with all galaxies (green line) and without gas-poor ($L_X < 10^{39}$ erg $s^{-1}$) galaxies (cyan line). The yellow line indicates the best fit of cD-type galaxies from O'Sullivan et al. (2003)

Figure 8. Similar to Figure 5, but with galaxies used in O'Sullivan et al. (2003) and David et al. (2006). Their results are compared with ours as indicated by arrows.

Figure 9. $L_X$(gas) is plotted against the average age of the stellar age for seven galaxies with similar $T_{gas}$ (0.32-0.36 keV) and similar $\sigma_*$. (202-232 km $s^{-1}$).

Figure A1. X-ray spectrum of the M32 diffuse emission with the best fit model. The blue line is for 0.5 keV APEC and the red line for power-law (PL) with $\Gamma$=1.8.

Figure A2. X-ray spectrum of the diffuse emission of the M31 bulge with the best fit model. The hot ISM dominates the diffuse emission at low energies (below 1-2 keV) and the stellar emission (ABs + CVs) dominates at high energies (above 2 keV). Spectral fitting is done iteratively, first fit the spectra at high energies with a power-law and then fit the spectrum in the entire energy range with APEC (gas) + APEC + PL. We allow the individual elements in APEC (gas) to vary independently. The blue line is for APEC (gas) and the red line for PL. The best fit normalization of APEC (AB+CV) is very low, so this component is not shown.

Figure A3. X-ray spectra of (a) M32 and (b) M31 with the best fit models determined by jointly fitting both spectra. The blue line is for APEC, the red line for PL and the green line for APEC (gas). All LMXBs are completely detected and removed in both galaxies.

Table 1
Early Type Galaxy Sample - optical properties

| name | T | $R_{25}$ (arcmin) | d (Mpc) | age (Gyr) | $\sigma_*$ (km/s) | B (mag) | $M_B$ (mag) | K (mag) | log $L_K$ | $S_N$ |
|---|---|---|---|---|---|---|---|---|---|---|
| N0221 | -6.0 | 4.3 | 0.8 | 2.4 | 72.1 | 8.72 | -15.82 | 5.09 | 9.1 | 1.00 |
| N0720 | -5.0 | 2.3 | 27.6 | 5.4 | 238.6 | 11.13 | -21.08 | 7.27 | 11.3 | 2.20 |
| N0821 | -5.0 | 1.2 | 24.1 | 8.9 | 188.7 | 11.72 | -20.19 | 7.90 | 10.9 | |
| N1023 | -3.0 | 4.3 | 11.4 | 4.7 | 210.0 | 10.08 | -20.21 | 6.23 | 10.9 | 0.00 |
| N1052 | -5.0 | 1.5 | 19.4 | 21.7 | 202.6 | 11.35 | -20.09 | 7.45 | 10.9 | 1.90 |
| N1316 | -2.0 | 6.0 | 21.4 | 3.2 | 223.1 | 9.40 | -22.26 | 5.58 | 11.7 | |
| N1427 | -4.1 | 1.8 | 23.5 | 12.2 | 171.0 | 11.81 | -20.05 | 8.14 | 10.8 | 4.20 |
| N1549 | -5.0 | 2.4 | 19.6 | 6.1 | 203.3 | 10.68 | -20.79 | 6.78 | 11.2 | 0.60 |
| N2434 | -5.0 | 1.2 | 21.5 | 5.5 | 180.4 | 11.57 | -20.10 | 7.88 | 10.8 | |
| N2768 | -5.0 | 4.0 | 22.3 | 10.0 | 211.0 | 10.70 | -21.05 | 6.99 | 11.2 | 0.00 |
| N3115 | -3.0 | 3.6 | 9.6 | 3.9 | 264.0 | 9.74 | -20.18 | 5.88 | 10.9 | 1.60 |
| N3377 | -5.0 | 2.6 | 11.2 | 3.6 | 107.6 | 11.07 | -19.18 | 7.44 | 10.4 | 2.40 |
| N3379 | -5.0 | 2.6 | 10.5 | 10.0 | 203.2 | 10.18 | -19.94 | 6.27 | 10.8 | 1.20 |
| N3384 | -3.0 | 2.7 | 11.5 | 3.2 | 170.0 | 10.75 | -19.57 | 6.75 | 10.7 | 0.90 |
| N3585 | -5.0 | 2.3 | 20.0 | 3.1 | 223.0 | 10.64 | -20.86 | 6.70 | 11.2 | |
| N3923 | -5.0 | 2.9 | 22.9 | 3.3 | 267.9 | 10.62 | -21.18 | 6.50 | 11.4 | 6.80 |
| N4125 | -5.0 | 2.8 | 23.8 | | 222.3 | 10.67 | -21.22 | 6.85 | 11.3 | |
| N4261 | -5.0 | 2.0 | 31.6 | 16.3 | 288.3 | 11.36 | -21.14 | 7.26 | 11.4 | |
| N4278 | -5.0 | 2.0 | 16.0 | 12.0 | 232.5 | 10.97 | -20.06 | 7.18 | 10.8 | 6.90 |
| N4365 | -5.0 | 3.4 | 20.4 | 5.9 | 270.0 | 10.49 | -21.06 | 6.64 | 11.2 | 3.86 |
| N4374 | -5.0 | 3.2 | 18.3 | 12.8 | 282.1 | 10.01 | -21.31 | 6.22 | 11.3 | 5.20 |
| N4382 | -1.0 | 3.5 | 18.4 | 1.6 | 189.0 | 9.99 | -21.34 | 6.14 | 11.4 | 1.29 |
| N4472 | -5.0 | 5.1 | 16.2 | 9.6 | 279.2 | 9.33 | -21.73 | 5.39 | 11.5 | 5.40 |
| N4473 | -5.0 | 2.2 | 15.7 | 4.0 | 201.0 | 11.03 | -19.94 | 7.15 | 10.8 | 1.98 |
| N4526 | -2.0 | 3.6 | 16.9 | 1.6 | 247.0 | 10.53 | -20.60 | 6.47 | 11.1 | 1.09 |
| N4552 | -5.0 | 2.5 | 15.3 | 12.4 | 251.8 | 10.57 | -20.36 | 6.72 | 11.0 | 2.82 |
| N4621 | -5.0 | 2.6 | 18.2 | 15.8 | 260.0 | 10.53 | -20.78 | 6.74 | 11.1 | 2.70 |
| N4649 | -5.0 | 3.7 | 16.8 | 14.1 | 309.8 | 9.70 | -21.43 | 5.73 | 11.4 | 5.16 |
| N4697 | -5.0 | 3.6 | 11.7 | 8.3 | 162.4 | 10.07 | -20.28 | 6.36 | 10.9 | 2.50 |
| N5866 | -1.0 | 2.3 | 15.3 | 1.8 | 175.0 | 10.83 | -20.10 | 6.87 | 10.9 | |

Table 2
Early Type Galaxy Sample - Chandra observations

| name | observation id(s) | obs date(s) | exposure (ksec) | $N_H$ ($10^{20}$ $cm^{-2}$) | limit* |
|---|---|---|---|---|---|
| N0221 | 313, 314, 1580, 2017, 2494, 5690 | Sep 21 2000 - May 27 2005 | 173 | 6.38 | 0.00** |
| N0720 | 492, 7372, 7062, 8448, 8449 | Oct 12 2000 - Oct 12 2006 | 127.77 | 1.57 | 1.31 |
| N0821 | 4006, 4408, 5692, 6310, 5691, 6313, 6314 | Nov 26 2002 - Jun 23 2005 | 208.91 | 6.23 | 0.37 |
| N1023 | 4696, 8198, 8464, 8465, 8197 | Feb 27 2004 - Jun 25 2007 | 194.60 | 7.05 | 0.11 |
| N1052 | 5910 | Sep 18 2005 | 57.42 | 3.10 | 0.89 |
| N1316 | 2022 | Apr 17 2001 | 24.09 | 2.13 | 1.97 |
| N1427 | 4742 | May 1 2005 | 50.25 | 1.33 | 0.92 |
| N1549 | 2077 | Nov 8 2000 - Sep 2 2001 | 25.38 | 1.48 | 1.28 |
| N2434 | 2923 | Oct 24 2002 | 24.39 | 12.23 | 1.86 |
| N2768 | 9528 | Jan 25 2008 | 63.22 | 4.11 | 1.05 |
| N3115 | 2040 | Jun 14 2001 | 34.75 | 4.61 | 0.23 |
| N3377 | 2934 | Jan 6 2003 | 39.25 | 2.77 | 0.21 |
| N3379 | 1587, 7073, 7074, 7075, 7076 | Feb 13 2001 - Jan 10 2007 | 324.21 | 2.78 | 0.06 |
| N3384 | 4692 | Oct 19 2004 | 9.90 | 2.74 | 0.63 |
| N3585 | 2078, 9506 | Jun 3 2001 - Mar 11 2008 | 90.17 | 5.60 | 0.55 |
| N3923 | 1563, 9507 | Jul 14 2001 - Apr 11 2008 | 93.42 | 6.30 | 0.94 |
| N4125 | 2071 | Sep 9 2001 | 60.68 | 1.82 | 1.28 |
| N4261 | 9569 | Feb 12 2008 | 98.77 | 1.58 | 1.93 |
| N4278 | 4741, 7077, 7078, 7079, 7080, 7081 | Feb 3 2005 - Feb 20 2007 | 457.98 | 1.76 | 0.14 |
| N4365 | 2015, 5921, 5922, 5923, 5924, 7224 | Jun 2 2001 - Nov 26 2005 | 190.67 | 1.61 | 0.32 |
| N4374 | 803 | May 19 2000 | 27.09 | 2.78 | 1.30 |
| N4382 | 2016 | May 29 2001 | 38.96 | 2.50 | 0.77 |
| N4472 | 321 | Jun 12 2000 | 32.08 | 1.62 | 1.39 |
| N4473 | 4688 | Feb 26 2005 | 29.58 | 2.65 | 0.69 |
| N4526 | 3925 | Nov 14 2003 | 38.20 | 1.63 | 0.63 |
| N4552 | 2072 | Apr 22 2001 | 47.90 | 2.56 | 1.02 |
| N4621 | 2068 | Aug 1 2001 | 23.06 | 2.17 | 0.85 |
| N4649 | 785, 8182, 8507 | Apr 20 2000 - Feb 1 2007 | 89.06 | 2.13 | 0.60 |
| N4697 | 4727, 4728, 4729, 4730 | Dec 26 2003 - Aug 18 2004 | 132.04 | 2.14 | 0.14 |
| N5866 | 2879 | Nov 14 2002 | 29.59 | 1.47 | 0.58 |

* 90% LMXB detection limit in unit of $10^{38}$ erg $s^{-1}$ in 0.3-8 keV.
** Minimum $L_x=9\times10^{33}$ erg $s^{-1}$

Table 3
Spectral fitting results

| name | region | Kfrac[a] | $\chi^2$/dof | T (keV) (1 σ error) | Γ (1 σ error) | $L_X$(LMXB/AGN) ($10^{40}$ erg $s^{-1}$) (1 σ error) | $L_X$(gas) ($10^{40}$ erg $s^{-1}$) (1 σ error) | $L_X$(APEC) ($10^{40}$ erg $s^{-1}$) | $L_X$(PL) ($10^{40}$ erg $s^{-1}$) |
|---|---|---|---|---|---|---|---|---|---|
| N0221 | AGN : | 0.06 | 29/ 29 | 1.00 | 2.36(-0.12+0.12) | 2.34e-04 (-1.2e-05 +1.2e-05) | 4.08e-24 (-4.1e-24 +3.2e-04) | 1.67e-05 | 5.54e-05 |
| N0221 | LMXB: | 0.05 | 245/275 | 1.00 | | 9.90e-03 (-1.0e-04 +1.0e-04) | 9.76e-05 (-3.9e-05 +3.9e-05) | 1.54e-05 | 5.11e-05 |
| N0221 | DIFF: | 0.66 | 106/138 | 1.00(-0.16+0.00) | | 7.85e-26 (-7.9e-26 +8.7e-04) | 1.98e-05 (-1.4e-05 +1.4e-05) | 1.89e-04 | 6.26e-04 |
| N0720 | AGN : | 0.05 | 10/ 16 | 0.54 | 1.16(-0.34+0.30) | 1.77e-01 (-2.7e-02 +3.2e-02) | 8.08e-02 (-1.1e-02 +1.1e-02) | 2.03e-03 | 6.75e-03 |
| N0720 | LMXB: | 0.15 | 108/140 | 0.54 | | 2.33e+00 (-6.3e-02 +6.3e-02) | 4.90e-01 (-2.5e-02 +2.5e-02) | 6.80e-03 | 2.26e-02 |
| N0720 | DIFF: | 0.77 | 231/216 | 0.54(-0.01+0.01) | | 4.63e-01 (-1.2e-01 +1.2e-01) | 4.51e+00 (-5.6e-02 +5.5e-02) | 3.45e-02 | 1.15e-01 |
| N0821 | AGN : | 0.08 | 3/ 5 | 0.15 | 1.58(-0.25+0.23) | 7.44e-02 (-8.7e-03 +1.0e-02) | 2.14e-03 (-2.1e-03 +0.0e+00) | 1.52e-03 | 5.05e-03 |
| N0821 | LMXB: | 0.22 | 55/ 73 | 0.15 | | 6.00e-01 (-6.0e-01 +3.5e-02) | 6.95e-23 ( 4.8e-15 +6.4e-01) | 4.25e-03 | 1.41e-02 |
| N0821 | DIFF: | 0.40 | 18/ 31 | 0.15(-0.05+0.85) | | 4.68e-02 (-4.7e-02 +4.6e-02) | 6.95e-23 (-7.0e-23 +9.3e-02) | 7.60e-03 | 2.52e-02 |
| N1023 | AGN : | 0.05 | 28/ 50 | 0.32 | 1.99(-0.06+0.06) | 1.10e-01 (-3.7e-03 +3.7e-03) | 2.16e-20 (-2.2e-20 +1.2e-01) | 9.44e-04 | 3.13e-03 |
| N1023 | LMXB: | 0.11 | 166/158 | 0.32 | | 3.65e-01 (-7.9e-03 +7.9e-03) | 1.32e-02 (-3.4e-03 +3.4e-03) | 2.11e-03 | 6.99e-03 |
| N1023 | DIFF: | 0.53 | 90/139 | 0.32(-0.01+0.02) | | 1.36e-02 (-7.0e-03 +7.0e-03) | 4.96e-02 (-3.6e-03 +3.6e-03) | 1.04e-02 | 3.47e-02 |
| N1052 | AGN : | 0.09 | 146/139 | 0.34 | -0.35(-0.05+0.05) | 1.16e+01 (-4.0e-01 +4.1e-01) | 1.40e-01 (-1.4e-02 +1.4e-02) | 1.76e-03 | 5.84e-03 |
| N1052 | LMXB: | 0.13 | 32/ 48 | 0.34 | | 8.19e-01 (-3.7e-02 +3.7e-02) | 5.17e-02 (-1.4e-02 +1.4e-02) | 2.36e-03 | 7.82e-03 |
| N1052 | DIFF: | 0.73 | 100/ 75 | 0.34(-0.02+0.02) | | 4.79e-01 (-4.2e-02 +4.2e-02) | 2.48e-01 (-2.0e-02 +2.0e-02) | 1.36e-02 | 4.50e-02 |
| N1316 | AGN : | 0.03 | 12/ 14 | 0.60 | 1.89(-0.26+0.22) | 3.85e-01 (-5.8e-02 +5.9e-02) | 1.97e-01 (-3.1e-02 +3.1e-02) | 3.58e-03 | 1.19e-02 |
| N1316 | LMXB: | 0.09 | 69/ 88 | 0.60 | | 3.12e+00 (-1.3e-01 +1.3e-01) | 9.58e-01 (-5.7e-02 +5.7e-02) | 1.16e-02 | 3.85e-02 |
| N1316 | DIFF: | 0.73 | 155/131 | 0.60(-0.01+0.01) | | 3.22e-01 (-2.8e-01 +2.8e-01) | 4.20e+00 (-1.2e-01 +1.2e-01) | 9.21e-02 | 3.05e-01 |
| N1427* | AGN : | 0.07 | 12/ 14 | 0.00 | 1.80( 0.00+0.00) | 2.00e-02 (-2.0e-02 +0.0e+00) | 0.00e+00 ( 0.0e+00 +0.0e+00) | 0.00e+00 | 0.00e+00 |
| N1427 | LMXB: | 0.17 | 31/ 36 | 0.38 | | 1.00e+00 (-1.0e+00 +5.0e-02) | 6.64e-23 (-6.6e-23 +1.1e+00) | 2.54e-03 | 8.41e-03 |
| N1427 | DIFF: | 0.79 | 50/ 65 | 0.38(-0.11+0.26) | | 1.74e-02 (-1.7e-02 +7.8e-02) | 5.94e-02 (-2.7e-02 +2.4e-02) | 1.14e-02 | 3.79e-02 |
| N1549 | AGN : | 0.05 | 0/ 1 | 0.35 | 1.90(-0.53+0.34) | 1.53e-01 (-2.8e-02 +2.6e-02) | 5.08e-03 (-5.1e-03 +2.0e-02) | 1.75e-03 | 5.79e-03 |
| N1549 | LMXB: | 0.10 | 14/ 28 | 0.35 | | 9.95e-01 (-1.0e+00 +6.0e-02) | 4.64e-23 (-4.6e-23 +1.1e+00) | 3.55e-03 | 1.18e-02 |
| N1549 | DIFF: | 0.87 | 61/ 86 | 0.35(-0.04+0.04) | | 2.66e-01 (-1.2e-01 +1.3e-01) | 3.05e-01 (-4.4e-02 +4.4e-02) | 3.08e-02 | 1.02e-01 |
| N2434* | AGN : | 0.06 | 0/ 1 | 0.00 | 1.80( 0.00+0.00) | 5.00e-02 (-5.0e-02 +0.0e+00) | 0.00e+00 ( 0.0e+00 +0.0e+00) | 0.00e+00 | 0.00e+00 |
| N2434 | LMXB: | 0.08 | 5/ 10 | 0.52 | | 6.40e-01 (-6.2e-02 +6.2e-02) | 4.03e-02 (-2.4e-02 +2.4e-02) | 1.26e-03 | 4.19e-03 |
| N2434 | DIFF: | 1.02 | 23/ 34 | 0.52(-0.05+0.04) | | 2.62e-21 (-2.6e-21 +8.4e-01) | 7.16e-01 (-5.1e-02 +5.1e-02) | 1.57e-02 | 5.20e-02 |
| N2768 | AGN : | 0.03 | 8/ 12 | 0.34 | 1.12(-0.21+0.21) | 4.91e-01 (-5.5e-02 +6.3e-02) | 1.67e-02 (-1.7e-02 +1.6e-02) | 1.22e-03 | 4.05e-03 |
| N2768 | LMXB: | 0.06 | 41/ 52 | 0.34 | | 1.10e+00 (-4.8e-02 +4.8e-02) | 5.59e-02 (-1.9e-02 +1.9e-02) | 2.26e-03 | 7.51e-03 |
| N2768 | DIFF: | 0.81 | 122/161 | 0.34(-0.01+0.01) | | 1.18e-20 (-1.2e-20 +1.4e+00) | 1.18e+00 (-4.9e-02 +4.9e-02) | 3.09e-02 | 1.02e-01 |

Table 3 (continued)

| gname | region | Kfrac | chi2/dof | T | gamma | Lx(LMXB/AGN) | Lx(gas) | Lx(APEC) | Lx(PL) |
|---|---|---|---|---|---|---|---|---|---|
| N3115 | AGN : | 0.04 | 2/ 4 | 0.61 | 1.68(-0.52+0.41) | 3.73e-02 (-6.4e-03 +6.5e-03) | 5.00e-03 (-4.1e-03 +3.8e-03) | 8.44e-04 | 2.80e-03 |
| N3115 | LMXB: | 0.13 | 53/ 62 | 0.61 | | 4.73e-01 (-4.7e-01 +2.4e-02) | 4.38e-23 (-4.4e-23 +5.0e-01) | 2.60e-03 | 8.62e-03 |
| N3115 | DIFF: | 0.38 | 23/ 36 | 0.61(-0.12+0.08) | | 5.15e-02 (-1.3e-02 +1.3e-02) | 2.66e-02 (-5.1e-03 +5.1e-03) | 1.14e-02 | 3.80e-02 |
| N3377 | AGN : | | | | | | | | |
| N3377 | LMXB: | 0.19 | 25/ 31 | 0.22 | | 2.73e-01 (-2.7e-01 +1.6e-02) | 1.51e-23 ( 3.7e-21 +2.9e-01) | 1.19e-03 | 3.93e-03 |
| N3377 | DIFF: | 0.79 | 31/ 54 | 0.22(-0.07+0.12) | | 1.29e-20 (-1.3e-20 +5.8e-02) | 1.18e-02 (-7.5e-03 +7.4e-03) | 4.93e-03 | 1.64e-02 |
| N3379 | AGN : | 0.04 | 19/ 27 | 0.25 | 1.92(-0.12+0.11) | 2.23e-02 (-1.4e-03 +1.4e-03) | 4.23e-04 (-4.2e-04 +9.2e-04) | 6.40e-04 | 2.12e-03 |
| N3379 | LMXB: | 0.20 | 273/255 | 0.25 | | 7.32e-01 (-6.7e-03 +6.8e-03) | 2.47e-02 (-2.6e-03 +2.7e-03) | 3.24e-03 | 1.08e-02 |
| N3379 | DIFF: | 0.71 | 235/317 | 0.25(-0.02+0.03) | | 6.70e-03 (-6.7e-03 +8.3e-03) | 2.20e-02 (-3.4e-03 +3.4e-03) | 1.16e-02 | 3.86e-02 |
| N3384* | AGN : | 0.05 | 19/ 27 | 0.00 | 1.80( 0.00+0.00) | 2.50e-02 (-2.5e-02 +0.0e+00) | 0.00e+00 ( 0.0e+00 +0.0e+00) | 0.00e+00 | 0.00e+00 |
| N3384 | LMXB: | 0.04 | 10/ 10 | 0.25 | | 4.34e-01 (-4.3e-01 +3.2e-02) | 1.61e-23 ( 1.4e+01 +4.7e-01) | 5.31e-04 | 1.76e-03 |
| N3384 | DIFF: | 0.86 | 12/ 19 | 0.25(-0.15+0.17) | | 8.55e-02 (-5.0e-02 +5.0e-02) | 3.50e-02 (-2.2e-02 +2.2e-02) | 1.09e-02 | 3.61e-02 |
| N3585 | AGN : | 0.06 | 9/ 12 | 0.36 | 1.84(-0.21+0.20) | 1.44e-01 (-1.4e-02 +1.4e-02) | 1.52e-02 (-9.4e-03 +9.0e-03) | 2.52e-03 | 8.35e-03 |
| N3585 | LMXB: | 0.07 | 46/ 65 | 0.36 | | 8.30e-01 (-3.0e-02 +3.1e-02) | 8.50e-03 (-8.5e-03 +1.1e-02) | 2.81e-03 | 9.34e-03 |
| N3585 | DIFF: | 0.82 | 96/141 | 0.36(-0.05+0.06) | | 1.41e-01 (-5.2e-02 +5.4e-02) | 1.23e-01 (-2.2e-02 +2.2e-02) | 3.23e-02 | 1.07e-01 |
| N3923 | AGN : | | | | | | | | |
| N3923 | LMXB: | 0.12 | 135/138 | 0.45 | | 2.35e+00 (-6.3e-02 +6.4e-02) | 9.01e-01 (-3.1e-02 +3.1e-02) | 7.42e-03 | 2.46e-02 |
| N3923 | DIFF: | 0.78 | 265/202 | 0.45(-0.01+0.01) | | 3.55e-01 (-1.1e-01 +1.1e-01) | 3.51e+00 (-5.6e-02 +5.6e-02) | 4.84e-02 | 1.61e-01 |
| N4125 | AGN : | 0.05 | 3/ 8 | 0.41 | 1.82(-0.46+0.35) | 1.47e-01 (-2.5e-02 +2.6e-02) | 5.48e-02 (-1.6e-02 +1.5e-02) | 2.44e-03 | 8.11e-03 |
| N4125 | LMXB: | 0.03 | 53/ 65 | 0.41 | | 1.40e+00 (-5.6e-02 +5.6e-02) | 1.61e-01 (-2.1e-02 +2.1e-02) | 1.65e-03 | 5.47e-03 |
| N4125 | DIFF: | 0.87 | 200/135 | 0.41(-0.01+0.01) | | 2.74e-20 (-2.7e-20 +3.2e+00) | 2.97e+00 (-4.7e-02 +4.8e-02) | 4.24e-02 | 1.41e-01 |
| N4261 | AGN : | 0.05 | 233/124 | 0.66 | -0.75(-0.10+0.10) | 9.15e+00 (-5.3e-01 +5.6e-01) | 2.15e+00 (-4.2e-02 +4.1e-02) | 3.23e-03 | 1.07e-02 |
| N4261 | LMXB: | 0.06 | 56/ 74 | 0.66 | | 1.72e+00 (-7.5e-02 +7.5e-02) | 1.64e-01 (-2.7e-02 +2.7e-02) | 3.59e-03 | 1.19e-02 |
| N4261 | DIFF: | 0.71 | 150/159 | 0.66(-0.01+0.01) | | 1.57e+00 (-1.4e-01 +1.4e-01) | 4.73e+00 (-7.5e-02 +7.4e-02) | 4.18e-02 | 1.39e-01 |
| N4278 | AGN : | 0.10 | 369/280 | 0.32 | 1.88(-0.01+0.01) | 2.19e+00 (-1.6e-02 +1.6e-02) | 1.03e-01 (-9.5e-03 +9.5e-03) | 1.57e-03 | 5.21e-03 |
| N4278 | LMXB: | 0.30 | 243/279 | 0.32 | | 1.27e+00 (-1.2e-02 +1.2e-02) | 4.63e-02 (-4.6e-03 +4.6e-03) | 4.85e-03 | 1.61e-02 |
| N4278 | DIFF: | 0.63 | 279/318 | 0.32(-0.01+0.01) | | 1.43e-01 (-2.0e-02 +2.0e-02) | 1.15e-01 (-7.8e-03 +7.8e-03) | 1.03e-02 | 3.42e-02 |
| N4365 | AGN : | 0.03 | 23/ 24 | 0.44 | 1.58(-0.12+0.11) | 1.56e-01 (-1.1e-02 +1.1e-02) | 2.89e-03 (-2.9e-03 +4.1e-03) | 1.34e-03 | 4.44e-03 |
| N4365 | LMXB: | 0.21 | 170/217 | 0.44 | | 2.68e+00 (-3.7e-02 +3.7e-02) | 7.39e-02 (-1.2e-02 +1.2e-02) | 9.09e-03 | 3.02e-02 |
| N4365 | DIFF: | 0.66 | 178/248 | 0.44(-0.02+0.02) | | 2.37e-01 (-5.0e-02 +5.0e-02) | 4.35e-01 (-1.8e-02 +1.8e-02) | 2.87e-02 | 9.52e-02 |

Table 3 (continued)

| gname | region | Kfrac | chi2/dof | T | gamma | Lx(LMXB/AGN) | Lx(gas) | Lx(APEC) | Lx(PL) |
|---|---|---|---|---|---|---|---|---|---|
| N4374 | AGN : | 0.04 | 24/ 31 | 0.63 | 1.40(-0.11+0.10) | 7.70e-01 (-6.4e-02 +6.6e-02) | 2.11e-01 (-2.4e-02 +2.4e-02) | 1.92e-03 | 6.36e-03 |
| N4374 | LMXB: | 0.08 | 133/ 96 | 0.63 | | 2.20e+00 (-8.9e-02 +8.8e-02) | 7.61e-01 (-3.8e-02 +3.8e-02) | 4.34e-03 | 1.44e-02 |
| N4374 | DIFF: | 0.80 | 128/141 | 0.63(-0.01+0.01) | | 3.74e-01 (-1.8e-01 +1.8e-01) | 5.00e+00 (-8.3e-02 +8.4e-02) | 4.12e-02 | 1.37e-01 |
| N4382 | AGN : | 0.03 | 24/ 31 | 0.00 | 0.00( 0.00+0.00) | 0.00e+00 ( 0.0e+00 +0.0e+00) | 0.00e+00 ( 0.0e+00 +0.0e+00) | 0.00e+00 | 0.00e+00 |
| N4382 | LMXB: | 0.05 | 39/ 64 | 0.40 | | 1.47e+00 (-5.5e-02 +5.5e-02) | 5.08e-02 (-1.9e-02 +1.9e-02) | 3.06e-03 | 1.02e-02 |
| N4382 | DIFF: | 0.80 | 116/115 | 0.40(-0.01+0.01) | | 5.65e-19 (-5.7e-19 +1.4e+00) | 1.14e+00 (-3.5e-02 +3.5e-02) | 4.47e-02 | 1.48e-01 |
| N4472 | AGN : | 0.02 | 51/ 37 | 0.80 | 3.83( Inf +Inf) | 4.84e-03 (-4.8e-03 +4.9e-01) | 4.04e-01 (-1.8e-02 +2.4e-02) | 1.38e-03 | 4.60e-03 |
| N4472 | LMXB: | 0.04 | 157/129 | 0.80 | | 2.33e+00 (-7.8e-02 +7.8e-02) | 6.44e-01 (-3.3e-02 +3.3e-02) | 3.20e-03 | 1.06e-02 |
| N4472 | DIFF: | 0.90 | 326/183 | 0.80(-0.00+0.00) | | 2.33e+00 (-2.6e-01 +2.6e-01) | 1.38e+01 (-1.6e-01 +1.6e-01) | 6.42e-02 | 2.13e-01 |
| N4473* | AGN : | 0.08 | 51/ 37 | 0.00 | 1.80( 0.00+0.00) | 2.00e-02 (-2.0e-02 +0.0e+00) | 0.00e+00 ( 0.0e+00 +0.0e+00) | 0.00e+00 | 0.00e+00 |
| N4473 | LMXB: | 0.04 | 11/ 18 | 0.35 | | 3.81e-01 (-2.9e-02 +2.9e-02) | 2.49e-02 (-1.1e-02 +1.1e-02) | 7.00e-04 | 2.32e-03 |
| N4473 | DIFF: | 0.85 | 54/ 76 | 0.35(-0.03+0.05) | | 1.80e-22 (-1.8e-22 +2.6e-01) | 1.60e-01 (-2.3e-02 +2.3e-02) | 1.35e-02 | 4.49e-02 |
| N4526 | AGN : | 0.04 | 3/ 7 | 0.33 | 1.07(-0.23+0.23) | 2.54e-01 (-3.2e-02 +3.6e-02) | 9.63e-03 (-9.0e-03 +8.4e-03) | 1.51e-03 | 5.02e-03 |
| N4526 | LMXB: | 0.11 | 34/ 41 | 0.33 | | 7.45e-01 (-3.6e-02 +3.7e-02) | 3.63e-02 (-1.4e-02 +1.4e-02) | 3.68e-03 | 1.22e-02 |
| N4526 | DIFF: | 0.71 | 58/ 69 | 0.33(-0.01+0.02) | | 1.34e-01 (-4.7e-02 +4.7e-02) | 2.82e-01 (-2.2e-02 +2.2e-02) | 2.47e-02 | 8.20e-02 |
| N4552 | AGN : | 0.07 | 83/ 53 | 0.52 | 1.70(-0.07+0.07) | 5.06e-01 (-2.7e-02 +2.7e-02) | 1.05e-01 (-1.2e-02 +1.2e-02) | 1.52e-03 | 5.06e-03 |
| N4552 | LMXB: | 0.09 | 98/121 | 0.52 | | 1.62e+00 (-4.5e-02 +4.5e-02) | 2.12e-01 (-1.7e-02 +1.7e-02) | 1.96e-03 | 6.52e-03 |
| N4552 | DIFF: | 0.81 | 124/153 | 0.52(-0.01+0.01) | | 3.67e-01 (-7.9e-02 +7.9e-02) | 2.00e+00 (-3.5e-02 +3.5e-02) | 1.83e-02 | 6.06e-02 |
| N4621 | AGN : | 0.07 | 0/ 1 | 0.27 | 1.85(-0.35+0.36) | 1.67e-01 (-2.5e-02 +3.1e-02) | 2.58e-19 (-2.6e-19 +2.0e-01) | 2.08e-03 | 6.91e-03 |
| N4621 | LMXB: | 0.10 | 23/ 24 | 0.27 | | 8.37e-01 (-8.4e-01 +5.5e-02) | 4.00e-23 ( 1.1e+06 +8.9e-01) | 3.22e-03 | 1.07e-02 |
| N4621 | DIFF: | 0.78 | 35/ 56 | 0.27(-0.09+0.13) | | 1.81e-01 (-9.9e-02 +9.7e-02) | 6.08e-02 (-3.7e-02 +3.7e-02) | 2.45e-02 | 8.14e-02 |
| N4649 | AGN : | 0.02 | 162/ 62 | 0.77 | 1.42(-0.18+0.15) | 1.27e-01 (-1.6e-02 +1.6e-02) | 3.66e-01 (-1.2e-02 +1.2e-02) | 1.49e-03 | 4.94e-03 |
| N4649 | LMXB: | 0.10 | 217/210 | 0.77 | | 3.38e+00 (-6.1e-02 +6.0e-02) | 1.48e+00 (-2.7e-02 +2.8e-02) | 6.91e-03 | 2.29e-02 |
| N4649 | DIFF: | 0.85 | 256/225 | 0.77(-0.00+0.00) | | 1.29e+00 (-1.7e-01 +1.7e-01) | 9.30e+00 (-1.1e-01 +1.1e-01) | 5.24e-02 | 1.74e-01 |
| N4697 | AGN : | 0.02 | 7/ 9 | 0.33 | 1.55(-0.13+0.13) | 3.22e-02 (-3.2e-03 +3.4e-03) | 1.65e-23 ( 5.3e-01 +3.6e-02) | 4.28e-04 | 1.42e-03 |
| N4697 | LMXB: | 0.15 | 145/180 | 0.33 | | 8.50e-01 (-1.4e-02 +1.4e-02) | 1.29e-02 (-5.1e-03 +5.0e-03) | 2.83e-03 | 9.40e-03 |
| N4697 | DIFF: | 0.72 | 170/202 | 0.33(-0.01+0.01) | | 1.28e-15 (-1.3e-15 +2.5e-01) | 1.78e-01 (-6.9e-03 +7.4e-03) | 1.33e-02 | 4.42e-02 |
| N5866* | AGN : | 0.04 | 7/ 9 | 0.00 | 1.80( 0.00+0.00) | 7.00e-02 (-7.0e-02 +0.0e+00) | 0.00e+00 ( 0.0e+00 +0.0e+00) | 0.00e+00 | 0.00e+00 |
| N5866 | LMXB: | 0.08 | 9/ 20 | 0.35 | | 3.71e-01 (-2.8e-02 +2.8e-02) | 3.11e-02 (-1.0e-02 +1.0e-02) | 1.55e-03 | 5.14e-03 |
| N5866 | DIFF: | 0.86 | 50/ 53 | 0.35(-0.02+0.03) | | 1.34e-01 (-4.4e-02 +4.4e-02) | 2.12e-01 (-1.9e-02 +1.9e-02) | 1.71e-02 | 5.67e-02 |

[a]Fraction of the K luminosity of the entire galaxy within the given region (AGN, LMXB, or DIFF)

*There was no AGN detected in N0224, N3377, and N3923. For the other galaxies flagged, we measured upper limits for the AGN luminosity by fixing the power law slope to 1.8 and subtracting a thermal spectrum scaled from the counts in the annulus surrounding the AGN region.

Table 4

Luminosity Ratio of Faint LMXBs

| $L_X$ | N3379 N | N3379 $L_X$ | N4278 N | N4278 $L_X$ | N4697 N | N4697 $L_X$ | combined |
|---|---|---|---|---|---|---|---|
| (Observed) | | | | | | | |
| 0-5 | 89 | 39.397 | 163 | 93.819 | 99 | 67.147 | |
| 1-5 | 9 | 20.772 | 25 | 47.737 | 19 | 35.600 | |
| 2-5 | 5 | 15.144 | 7 | 22.549 | 5 | 15.846 | |
| 0.1-1 | 50 | 16.805 | 125 | 45.098 | 74 | 31.062 | |
| 0.01-0.1 | 30 | 1.820 | 13 | 0.984 | 6 | 0.485 | |
| (before correction for undetected faint LMXBs) | | | | | | | |
| $R_{15}$ | | 1.90 | | 1.97 | | 1.89 | |
| $R_{25}$ | | 2.60 | | 4.16 | | 4.24 | |
| (expected $L_X$ by scaling the M32 value based on the $L_K$ ratio) | | | | | | | |
| 0-0.1 | | 1.905 | | 1.905 | | 2.398 | |
| (expected $L_X$ from undetected LMXBs) | | | | | | | |
| 0-0.1 | | 0.084 | | 0.921 | | 1.913 | |
| (after correction for undetected faint LMXBs) | | | | | | | |
| $R_{15}$ | | 1.90 | | 1.98 | | 1.94 | 1.95+/-0.04 |
| $R_{25}$ | | 2.61 | | 4.20 | | 4.36 | 3.80+/-0.97 |

- All $L_X$ are in unit of $10^{38}$ erg $s^{-1}$.
- $R_{15}$ and $R_{25}$ are defined in Section 3.4.2

Table 5

Summary of X-ray luminosity from individual components

| Name | $L_X$(gas) ±1σ error ($10^{40}$ erg $s^{-1}$) | $L_X$(LMXB) ±1σ ($10^{40}$ erg $s^{-1}$) | $L_X$(AGN) ±1σ ($10^{40}$ erg $s^{-1}$) | $L_X$(AB) | $L_X$(CV) | $L_X$(total) ±1σ |
|---|---|---|---|---|---|---|
| N0221 | 1.17e-04(-4.1e-05 +4.1e-05) | 9.90e-03(-1.0e-04 +1.0e-04) | 2.34e-04(-1.2e-05 +1.2e-05) | 0.00 | 0.00 | 0.01(-0.00 +0.00) |
| N0224 | 1.18e-02(-7.0e-04 +7.0e-04) | 5.02e-01(-2.1e-03 +2.1e-03) | *0.00e+00( 0.0e+00 +0.0e+00) | 0.00 | 0.00 | 0.52(-0.00 +0.00) |
| N0720 | 5.06e+00(-6.2e-02 +6.2e-02) | 2.80e+00(-1.3e-01 +1.3e-01) | 1.77e-01(-2.7e-02 +3.2e-02) | 0.04 | 0.14 | 8.22(-0.15 +0.15) |
| N0821 | 2.13e-03(-2.1e-03 +9.9e-03) | 6.47e-01(-6.0e-01 +5.8e-02) | 7.44e-02(-8.7e-03 +1.0e-02) | 0.01 | 0.04 | 0.78(-0.60 +0.06) |
| N1023 | 6.25e-02(-4.9e-03 +4.9e-03) | 3.79e-01(-1.1e-02 +1.1e-02) | 1.10e-01(-3.7e-03 +3.7e-03) | 0.01 | 0.04 | 0.61(-0.01 +0.01) |
| N1052 | 4.37e-01(-2.8e-02 +2.8e-02) | 1.30e+00(-5.6e-02 +5.6e-02) | 1.16e+01(-4.0e-01 +4.1e-01) | 0.02 | 0.06 | 13.41(-0.41 +0.42) |
| N1316 | 5.35e+00(-1.4e-01 +1.4e-01) | 3.44e+00(-3.1e-01 +3.1e-01) | 3.85e-01(-5.8e-02 +5.9e-02) | 0.11 | 0.36 | 9.64(-0.34 +0.34) |
| N1427 | 5.94e-02(-2.7e-02 +2.4e-02) | 1.02e+00(-1.0e+00 +9.3e-02) | *2.00e-02(-2.0e-02 +0.0e+00) | 0.01 | 0.05 | 1.14(-1.00 +0.10) |
| N1549 | 3.08e-01(-4.4e-02 +4.8e-02) | 1.26e+00(-1.0e+00 +1.5e-01) | 1.53e-01(-2.8e-02 +2.6e-02) | 0.04 | 0.12 | 1.88(-1.00 +0.16) |
| N2434 | 7.56e-01(-5.6e-02 +5.6e-02) | 6.40e-01(-6.2e-02 +6.2e-02) | *5.00e-02(-5.0e-02 +0.0e+00) | 0.02 | 0.06 | 1.47(-0.08 +0.08) |
| N2768 | 1.26e+00(-5.5e-02 +5.5e-02) | 1.10e+00(-4.8e-02 +4.8e-02) | 4.91e-01(-5.5e-02 +6.3e-02) | 0.03 | 0.11 | 3.00(-0.09 +0.10) |
| N3115 | 2.59e-02(-5.6e-03 +5.3e-03) | 5.21e-01(-4.7e-01 +2.6e-02) | 3.73e-02(-6.4e-03 +6.5e-03) | 0.01 | 0.04 | 0.63(-0.47 +0.03) |
| N3377 | 1.17e-02(-7.2e-03 +7.4e-03) | 2.73e-01(-2.7e-01 +1.6e-02) | *0.00e+00( 0.0e+00 +0.0e+00) | 0.01 | 0.02 | 0.31(-0.27 +0.02) |
| N3379 | 4.69e-02(-4.3e-03 +4.4e-03) | 7.38e-01(-9.5e-03 +1.1e-02) | 2.23e-02(-1.4e-03 +1.4e-03) | 0.02 | 0.05 | 0.87(-0.01 +0.01) |
| N3384 | 3.50e-02(-2.2e-02 +2.2e-02) | 5.19e-01(-4.4e-01 +5.9e-02) | *2.50e+00(-2.5e-02 +0.0e+00) | 0.01 | 0.04 | 0.60(-0.44 +0.06) |
| N3585 | 1.47e-01(-2.5e-02 +2.6e-02) | 9.71e-01(-6.0e-02 +6.2e-02) | 1.44e-01(-1.4e-02 +1.4e-02) | 0.04 | 0.12 | 1.42(-0.07 +0.07) |
| N3923 | 4.41e+00(-6.4e-02 +6.4e-02) | 2.71e+00(-1.3e-01 +1.3e-01) | *0.00e+00( 0.0e+00 +0.0e+00) | 0.06 | 0.19 | 7.36(-0.15 +0.15) |
| N4125 | 3.18e+00(-5.4e-02 +5.5e-02) | 1.40e+00(-5.6e-02 +5.6e-02) | 1.47e-01(-2.5e-02 +2.6e-02) | 0.05 | 0.15 | 4.92(-0.08 +0.08) |
| N4261 | 7.02e+00(-8.9e-02 +8.9e-02) | 3.30e+00(-1.6e-01 +1.6e-01) | 9.15e+00(-5.3e-01 +5.6e-01) | 0.05 | 0.16 | 19.68(-0.56 +0.58) |
| N4278 | 2.63e-01(-1.3e-02 +1.3e-02) | 1.42e+00(-2.3e-02 +2.3e-02) | 2.19e+00(-1.6e-02 +1.6e-02) | 0.02 | 0.06 | 3.94(-0.03 +0.03) |
| N4365 | 5.12e-01(-2.2e-02 +2.2e-02) | 2.91e+00(-6.3e-02 +6.3e-02) | 1.56e-01(-1.1e-02 +1.1e-02) | 0.04 | 0.13 | 3.75(-0.07 +0.07) |
| N4374 | 5.95e+00(-9.4e-02 +9.5e-02) | 2.57e+00(-2.0e-01 +2.0e-01) | 7.70e-01(-6.4e-02 +6.6e-02) | 0.05 | 0.16 | 9.50(-0.23 +0.23) |
| N4382 | 1.19e+00(-4.0e-02 +4.0e-02) | 1.47e+00(-5.5e-02 +5.5e-02) | 7.00e-02(-2.0e-02 +2.0e-02) | 0.05 | 0.16 | 2.86(-0.07 +0.07) |
| N4472 | 1.89e+01(-2.5e-01 +2.5e-01) | 9.45e+00(-4.2e-01 +4.2e-01) | 4.87e-03(-4.9e-03 +3.3e-02) | 0.08 | 0.28 | 28.70(-0.49 +0.48) |
| N4473 | 1.85e-01(-2.5e-02 +2.5e-02) | 3.81e-01(-2.9e-02 +2.9e-02) | 0.00e+00( 0.0e+00 +0.0e+00) | 0.01 | 0.05 | 0.63(-0.04 +0.04) |
| N4526 | 3.28e-01(-2.7e-02 +2.7e-02) | 8.79e-01(-5.9e-02 +5.9e-02) | 2.54e-01(-3.2e-02 +3.6e-02) | 0.03 | 0.10 | 1.59(-0.07 +0.07) |
| N4552 | 2.31e+00(-4.0e-02 +4.0e-02) | 1.99e+00(-9.1e-02 +9.1e-02) | 5.06e-01(-2.7e-02 +2.7e-02) | 0.02 | 0.07 | 4.90(-0.10 +0.10) |
| N4621 | 6.08e-02(-3.7e-02 +3.7e-02) | 1.02e+00(-8.4e-01 +1.1e-01) | 1.67e-01(-2.5e-02 +3.1e-02) | 0.03 | 0.10 | 1.37(-0.84 +0.12) |
| N4649 | 1.17e+01(-1.8e-01 +1.8e-01) | 5.04e+00(-3.1e-01 +3.1e-01) | 1.27e-01(-1.6e-02 +1.6e-02) | 0.07 | 0.22 | 17.17(-0.36 +0.36) |
| N4697 | 1.91e-01(-8.6e-03 +8.9e-03) | 8.50e-01(-1.4e-02 +1.4e-02) | 3.22e-02(-3.2e-03 +3.4e-03) | 0.02 | 0.06 | 1.14(-0.02 +0.02) |
| N5866 | 2.42e-01(-2.2e-02 +2.2e-02) | 5.04e-01(-5.2e-02 +5.2e-02) | *7.00e-02(-7.0e-02 +0.0e+00) | 0.02 | 0.06 | 0.83(-0.06 +0.06) |

* For N0224, N3377, and N3923 there is no AGN detectable. For the other galaxies flagged, we have provided upper limits by subtracting the expected thermal contribution in the AGN region by scaling from the count rate at a surrounding annulus

Table 6

Partial Correlation Coefficients

| | $L_X>10^{38}$ erg $s^{-1}$ | $L_X>10^{39}$ | $L_X>5x10^{39}$ | $L_X<5x10^{39}$ | $\sigma>240$ km $s^{-1}$ | $240>\sigma>200$ | $200>\sigma$ |
|---|---|---|---|---|---|---|---|
| (Lk,Lx) | | | | | | | |
| Spearman $r_s$ | 0.82 | 0.76 | | | 0.91 | 0.69 | 0.94 |
| Probability | $10^{-7}$ | $7x10^{-5}$ | | | $2x10^{-4}$ | 0.02 | $2x10^{-3}$ |
| ($\sigma$,Lx) | | | | | | | |
| Spearman $r_s$ | 0.62 | 0.69 | | | 0.64 | 0.63 | 0.90 |
| Probability | $5x10^{-4}$ | $5x10^{-4}$ | | | 0.04 | 0.04 | $5x10^{-3}$ |
| (kT,Lx) | | | | | | | |
| Spearman $r_s$ | 0.82 | 0.89 | | | 0.92 | 0.81 | 0.72 |
| Probability | $10^{-7}$ | $6x10^{-8}$ | | | $10^{-4}$ | $2x10^{-3}$ | 0.07 |
| (Lk,kT) | | | | | | | |
| Spearman $r_s$ | 0.71 | 0.67 | | | 0.81 | 0.70 | 0.61 |
| Probability | $2x10^{-5}$ | $8x10^{-4}$ | | | $4x10^{-3}$ | 0.02 | 0.1 |
| ($\sigma$,kT) | | | | | | | |
| Spearman $r_s$ | 0.63 | 0.60 | 0.78 | 0.16 | 0.67 | 0.65 | 0.91 |
| Probability | $3x10^{-4}$ | $4x10^{-3}$ | $2x10^{-3}$ | 0.58 | 0.03 | 0.03 | $5x10^{-3}$ |
| ($\sigma$,Lk) | | | | | | | |
| Spearman $r_s$ | 0.73 | 0.64 | | | 0.80 | 0.60 | 0.77 |
| Probability | $9x10^{-6}$ | $2x10^{-3}$ | | | $6x10^{-3}$ | 0.05 | 0.04 |

Table A1

Chandra observations used to determine AB and CV spectral parameters

| name | Obsid | Exposure (ksec) | $N_H$ ($10^{20}$ $cm^{-2}$) | d (Mpc) | Radius[a] (″) | $K_{tot}$ (mag) | $K_{diffuse}$[b] (mag) |
|---|---|---|---|---|---|---|---|
| N0221 | 313,314,1580,2017,5690,2494 | 173 | 6.38 | 0.821 | 60 | 5.096 | 5.539 |
| N0224 | 309,310,1854,1575 | 49 | 6.68 | 0.760 | 60 | 0.984 | 3.568 |
| N0821 | 4006,4408,5692,6310,5691,6313,6314 | 206 | 6.20 | 24.10 | 30 | 7.90 | 8.894 |
| N3379 | 1587,7073,7074,7075,7076 | 324 | 2.80 | 10.57 | 90 | 6.27 | 6.752 |

[a]The region in which diffuse emission was extracted

[b]K magnitude within the region of diffuse emission

Table A2

Spectral parameters of ABs and CVs in individual fitting

| name | $\chi^2_\nu,\nu$ | $L_X/L_K$ [a] (PL) | $\Gamma$ (PL) | $L_X/L_K$ [a] (APEC) | kT (APEC) | Z (APEC) | $L_X/L_K$ [a] (ISM) |
|---|---|---|---|---|---|---|---|
| N0221 (M32) | 0.789,136 | $7.9^{+1.0}_{-1.7}$ | $1.8^{+0.2}_{-0.6}$ | $2.0^{+2.2}_{-1.0}$ | $0.49^{+0.14}_{-0.07}$ | 0.02(>0.05) | 0 |
| ISM abundances linked at Solar ratios | | | | | | | |
| N0224 (M31) | 1.16,136 | $6.0^{+1.4}_{-0.7}$ | 1.5±0.4 | 8.9±0.5 | 0.55±0.01 | 5 (>2) | 10.9±0.5 |
| ISM abundances independent | | | | | | | |
| N0224 (M31) | 0.828,126 | $7.1^{+2.4}_{-1.2}$ | 1.7±0.4 | 0.00 | 0.5 [fixed] | ... | 17.4±0.3 |

[a] The 0.3-8.0 keV model luminosity divided by the K band luminosity in the diffuse region, in units of $10^{27}$ erg $s^{-1}/L_{K\odot}$

Table A3
Spectral parameters of ABs and CVs in joint fitting

| name | $\chi^2_\nu,\nu$ | $L_X/L_K$ [a] (PL) | PL $\Gamma$ (PL) | $L_X/L_K$ [a] (APEC) | kT (APEC) | Z (APEC) | $L_X/L_K$ [a] (ISM) |
|---|---|---|---|---|---|---|---|
| ISM abundances linked | | | | | | | |
| N0221 | 1.18,277 | $6.9^{+1.9}_{-1.0}$ | $1.8^{+0.3}_{-0.1}$ | $1.8^{+0.3}_{-0.1}$ | $0.60\pm0.03$ | 5(>0.7) | 0 |
| N0224 | | | | | | | $16.4^{+0.2}_{-0.4}$ |
| ISM abundances independent | | | | | | | |
| N0221 | 0.808,266 | $7.2^{+2.1}_{-1.1}$ | $1.76\pm0.37$ | $2.2\pm0.3$ | $0.48^{+0.07}_{-0.05}$ | $0.18^{+0.19}_{-0.07}$ | 0 |
| N0224 | | | | | | | $14.6^{+0.8}_{-0.4}$ |

[a] The 0.3-8.0 keV model luminosity divided by the K band luminosity in the diffuse region, in units of $10^{27}$ erg $s^{-1}/L_{K\odot}$

Table A4
$L_X/L_K$ from ABs and CVs

| Energy Range (keV) | 0.3-0.7 | 0.3-2 | 0.3-5 | 0.3-8 | 0.5-2 | 0.5-8 | 2-5 | 2-7 | 2-8 | 2-10 |
|---|---|---|---|---|---|---|---|---|---|---|
| Bandpass Name | ... | ... | ... | B | Sc | Bc | ... | ... | Hc | ... |
| Total $L_X/L_K$ | $2.3^{+1.4}_{-0.7}$ | $5.7^{+2.4}_{-1.4}$ | $8.0^{+2/4}_{-1.4}$ | $9.5^{+2.1}_{-1.1}$ | $4.4^{+1.5}_{-0.9}$ | $8.2^{+1.8}_{-0.9}$ | $2.4^{+1.7}_{-1.8}$ | 3.39±0.33 | $3.8^{+0.8}_{-0.4}$ | $4.5^{+0.8}_{-0.6}$ |
| Power Law | $1.4^{+1.4}_{-0.7}$ | $3.5^{+2.4}_{-1.4}$ | $5.8^{+2.4}_{-1.4}$ | $7.3^{+2.1}_{-1.1}$ | $2.7^{+1.5}_{-0.9}$ | $6.5^{+1.9}_{-1.0}$ | $2.4^{+0.1}_{-0.2}$ | 3.35±0.33 | $3.8^{+1.1}_{-0.6}$ | $4.5^{+0.8}_{-0.6}$ |
| APEC | 0.9±0.2 | 2.2±0.3 | 2.2±0.1 | 2.2±0.1 | $1.7^{+0.1}_{-0.2}$ | 1.7±0.1 | $0.04^{+0.01}_{-0.02}$ | $0.04^{+0.01}_{-0.02}$ | $0.04^{+0.01}_{-0.02}$ | $0.04^{+0.01}_{-0.02}$ |
| Rev. et al. 2007a, NGC 221 | | | | | 4.1±0.6 | | | 2.7±0.8 | | 3.3±1.0 |
| Li & Wang 2007, NGC 221 | | | | | 5.8±1.1 | | | | | 5.6±1.1 |
| Rev. et al. 2008 (all galaxies) | | | | | 5.9±2.5 | | | | | |
| Rev. et al. 2008 (NGC 3379) | | | | | 6.9±0.7 | | | | | |
| Bogdán & Gilfanov 2010 | 2.4±0.4 | | | | | | | | | |

$L_X/L_K$ in units of $10^{27}$ erg $s^{-1}/L_{K\odot}$

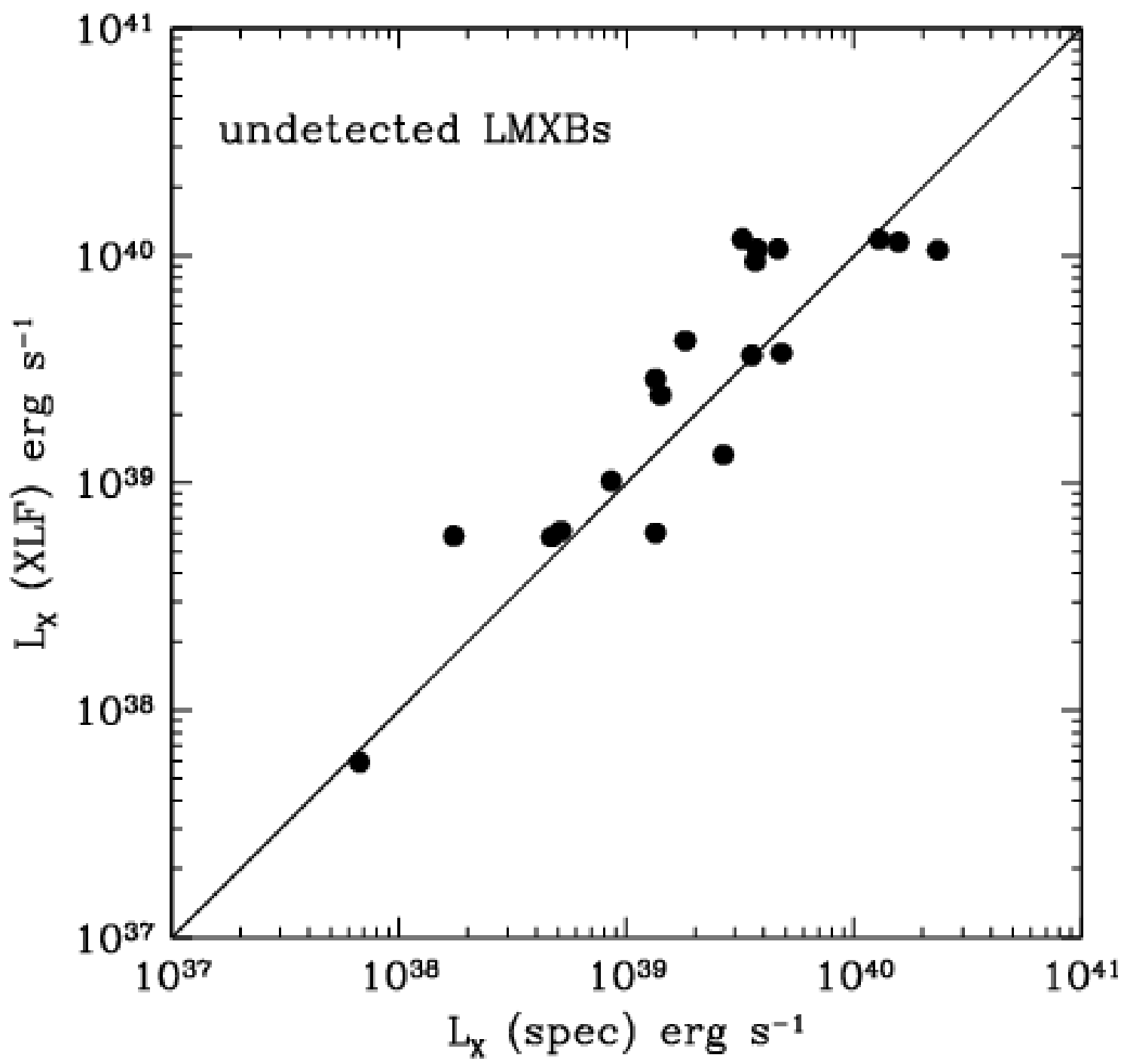
undetected LMXBs
$L_X$ (XLF) erg s$^{-1}$
$L_X$ (spec) erg s$^{-1}$
$10^{37}$
$10^{38}$
$10^{39}$
$10^{40}$
$10^{41}$

Figure 1

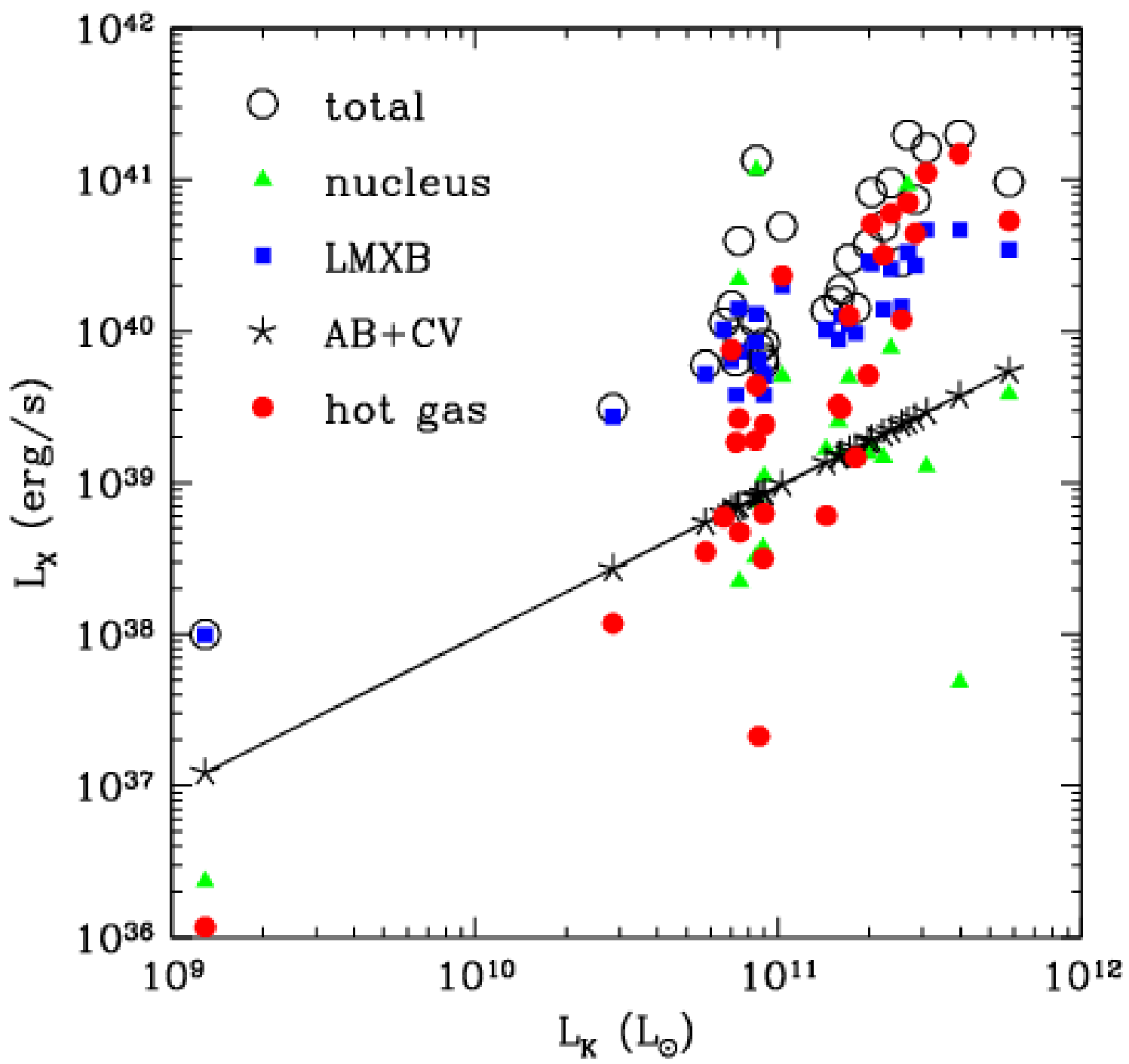
total
nucleus
LMXB
AB+CV
hot gas
$L_X$ (erg/s)
$L_K$ ($L_\odot$)
$10^{42}$
$10^{41}$
$10^{40}$
$10^{39}$
$10^{38}$
$10^{37}$
$10^{36}$
$10^{9}$
$10^{10}$
$10^{11}$
$10^{12}$

Figure 2

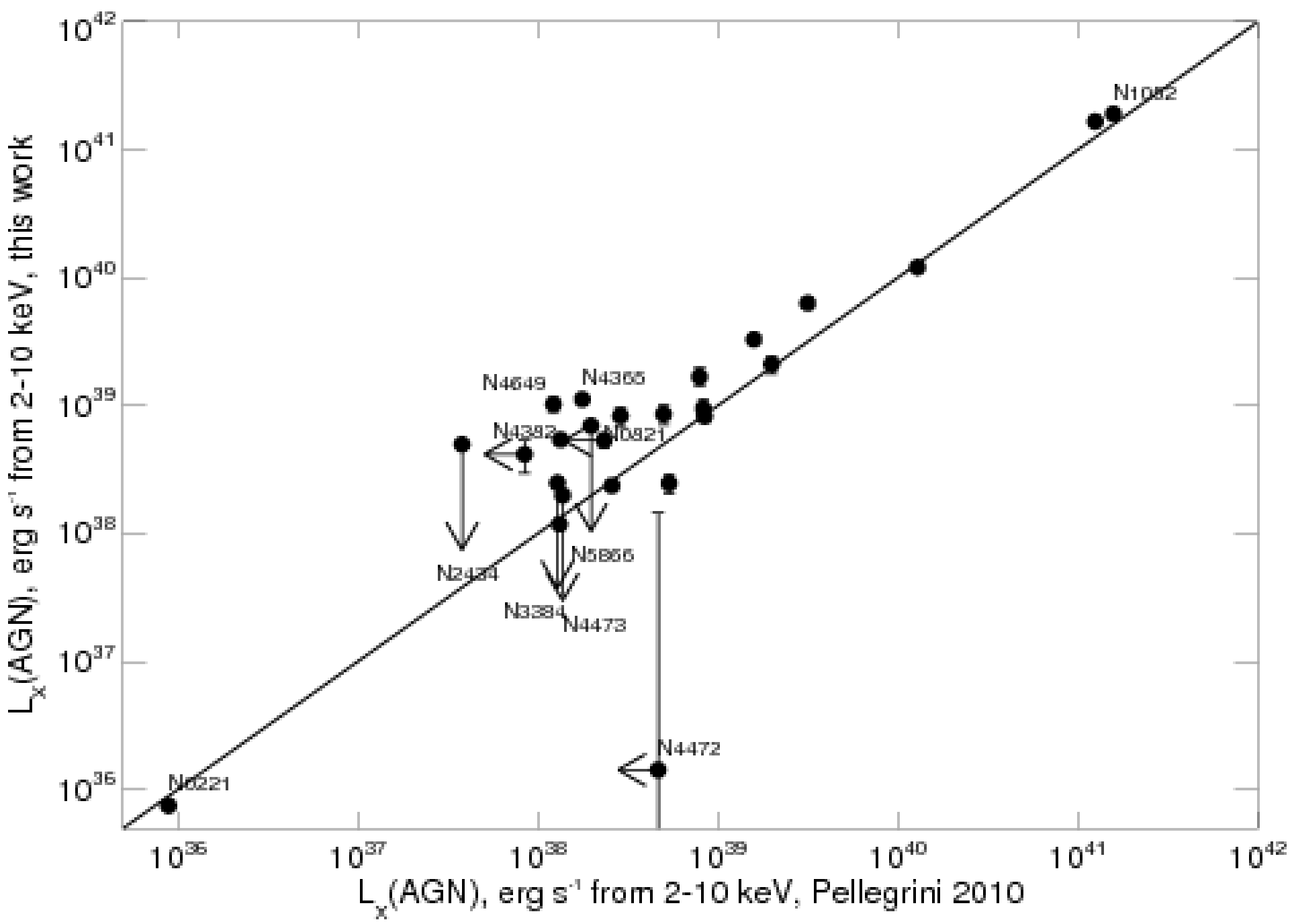
N1052
N4649
N4365
N4382
N0821
N2434
N5866
N3384
N4473
N4472
N0221
L_x(AGN), erg s^-1 from 2-10 keV, this work
L_x(AGN), erg s^-1 from 2-10 keV, Pellegrini 2010

Figure 3

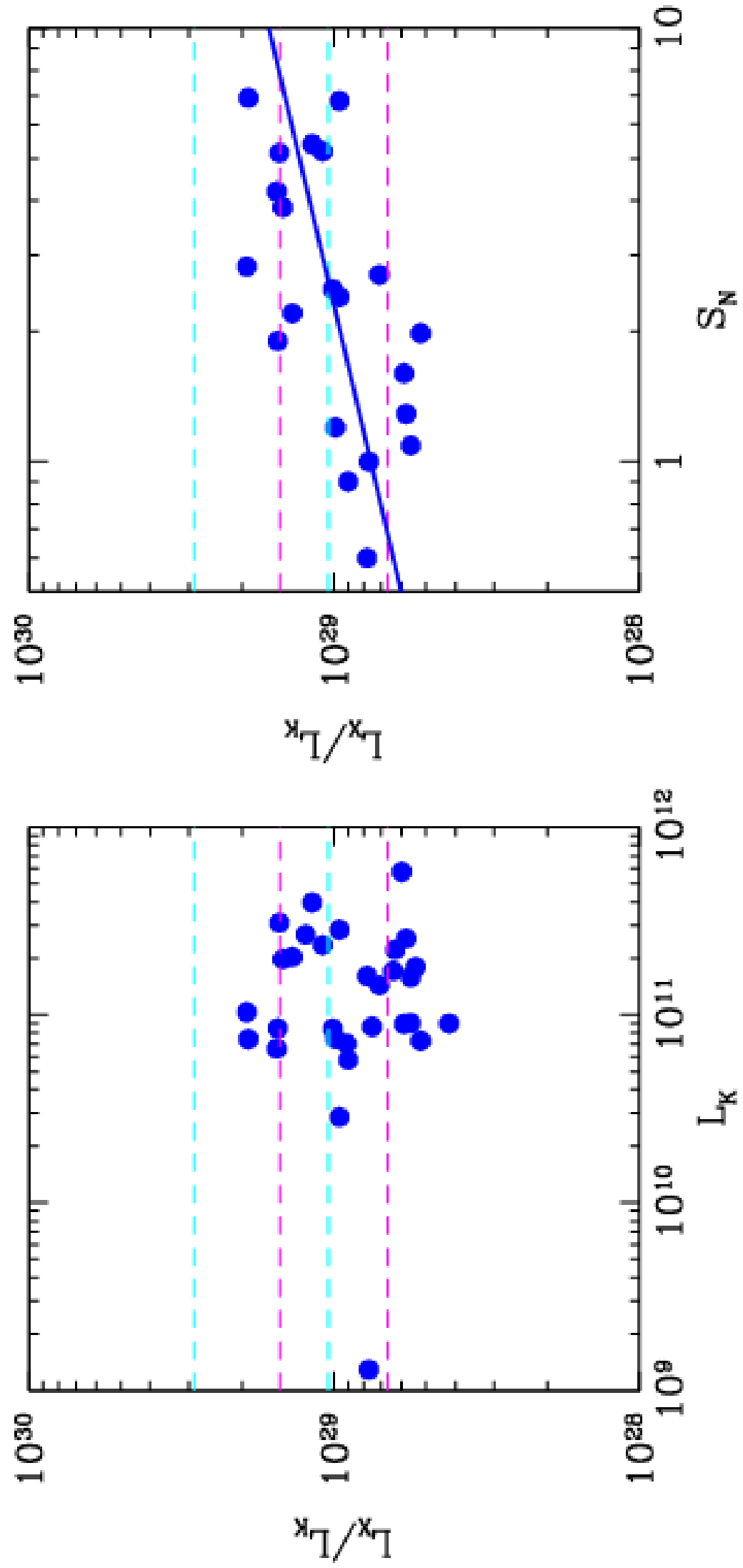

Figure 4

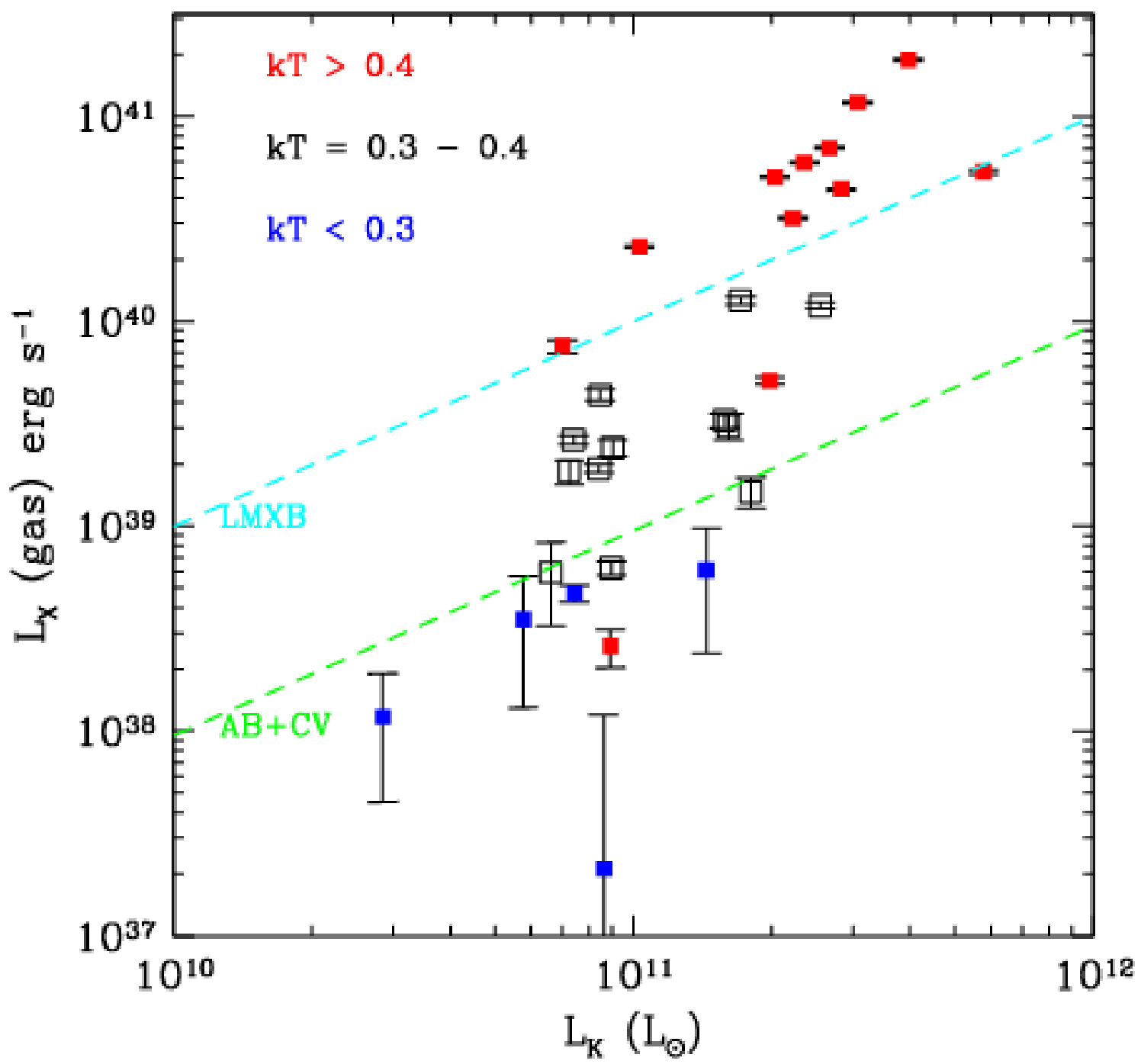
kT > 0.4
kT = 0.3 − 0.4
kT < 0.3
LMXB
AB+CV
$L_X$ (gas) erg $s^{-1}$
$L_K$ ($L_\odot$)
$10^{37}$
$10^{38}$
$10^{39}$
$10^{40}$
$10^{41}$
$10^{10}$
$10^{11}$
$10^{12}$

Figure 5a

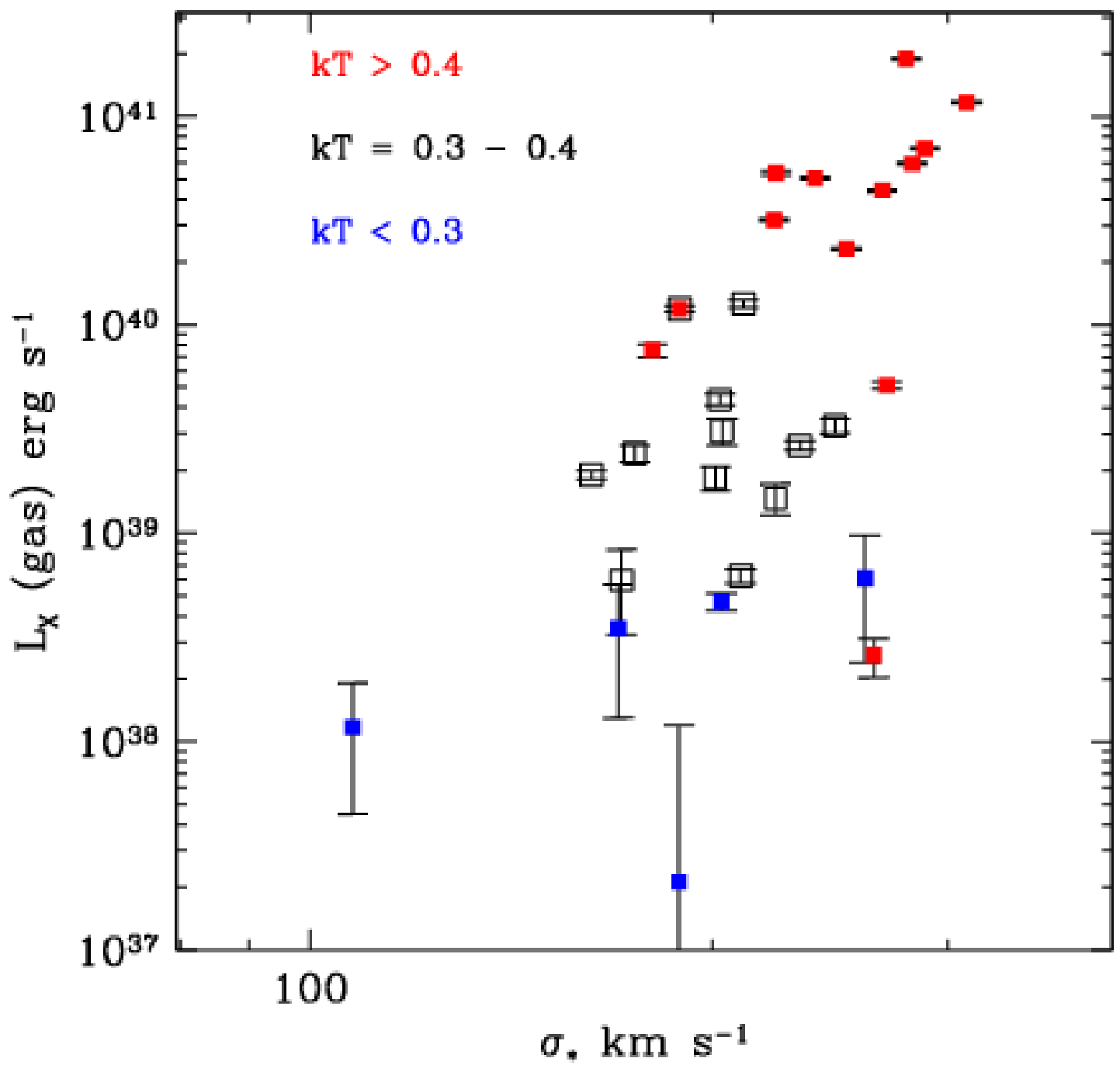

kT > 0.4
kT = 0.3 − 0.4
kT < 0.3
10^41
10^40
10^39
10^38
10^37
100
L_X (gas) erg s^-1
σ, km s^-1

Figure 5b

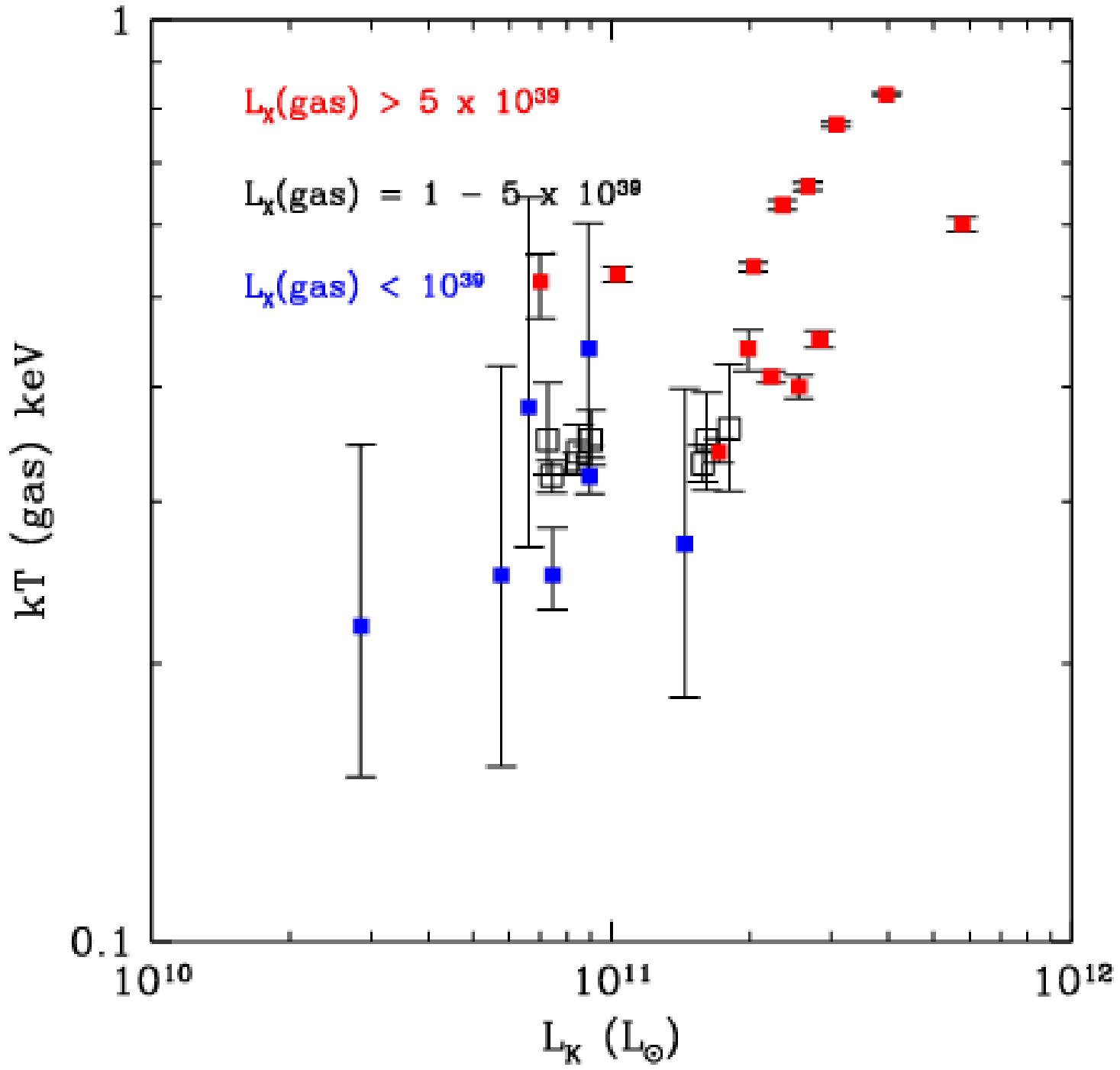

1
0.1
$10^{10}$
$10^{11}$
$10^{12}$
kT (gas) keV
$L_K$ ($L_\odot$)
$L_X$(gas) > 5 x $10^{39}$
$L_X$(gas) = 1 – 5 x $10^{39}$
$L_X$(gas) < $10^{39}$

Figure 6a

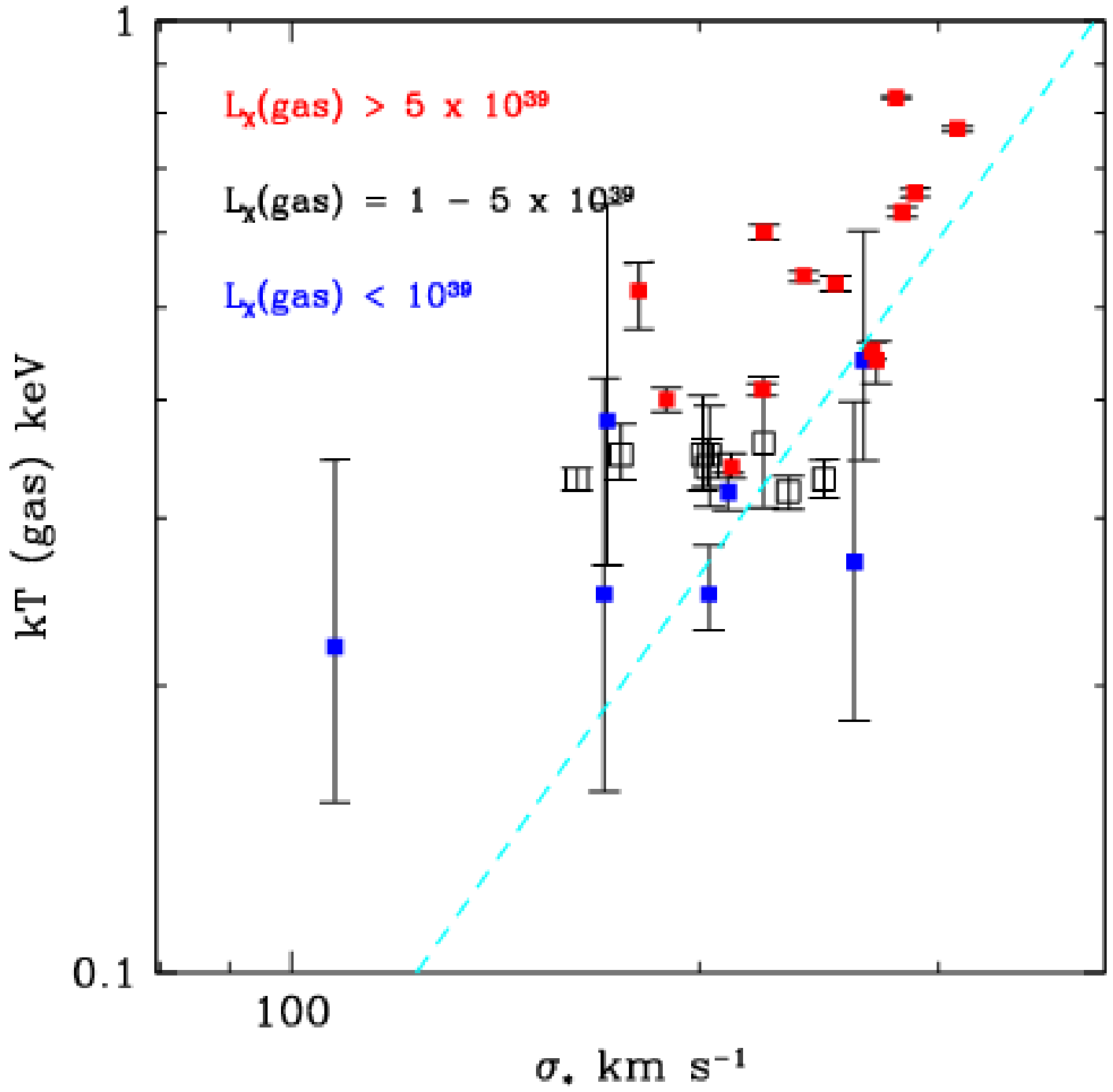

1
0.1
100
kT (gas) keV
σ* km s⁻¹
Lx(gas) > 5 x 10³⁹
Lx(gas) = 1 − 5 x 10³⁹
Lx(gas) < 10³⁹

Figure 6b

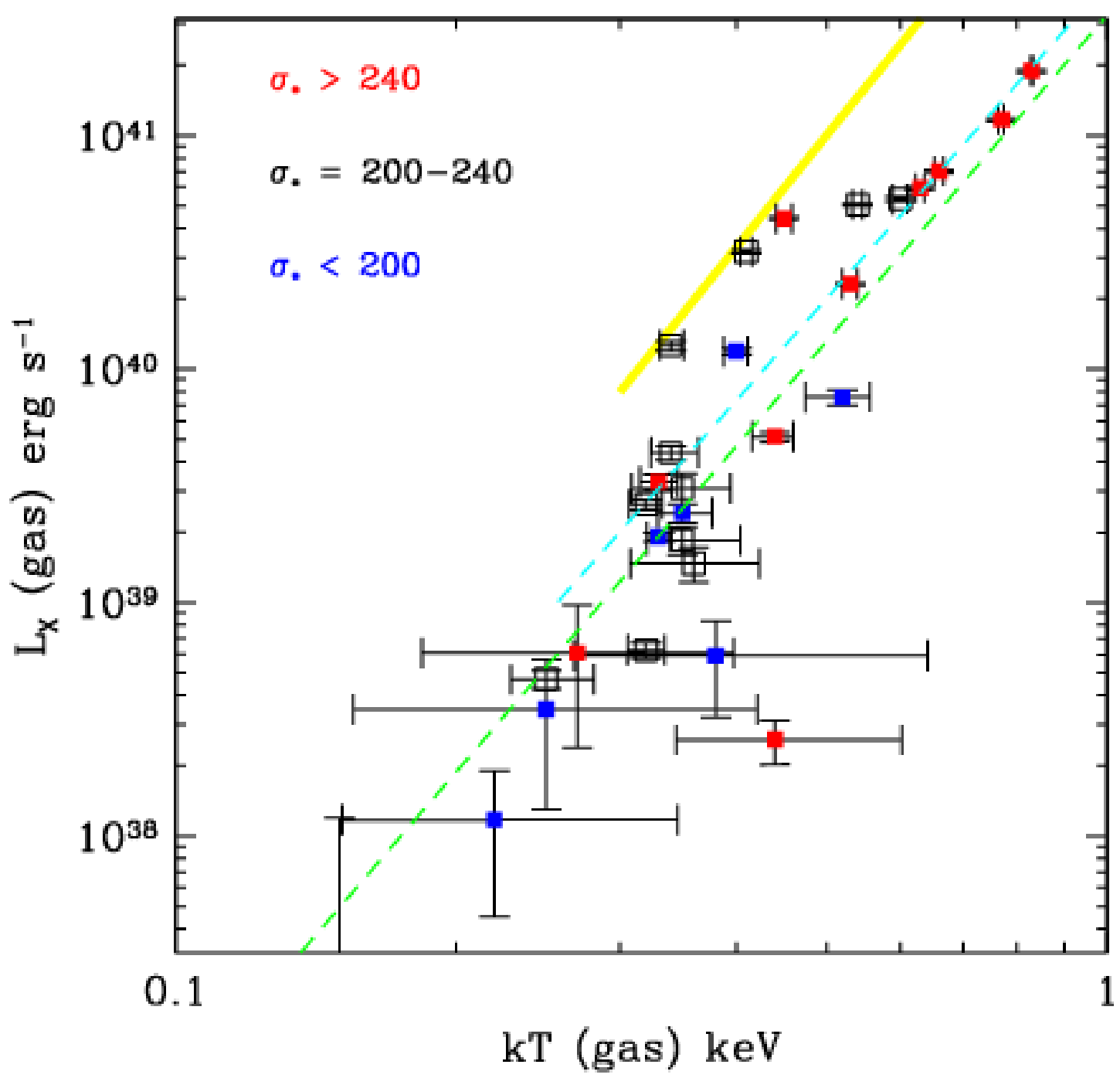
σ* > 240
σ* = 200−240
σ* < 200
10^41
10^40
10^39
10^38
L_X (gas) erg s^-1
0.1
1
kT (gas) keV

Figure 7

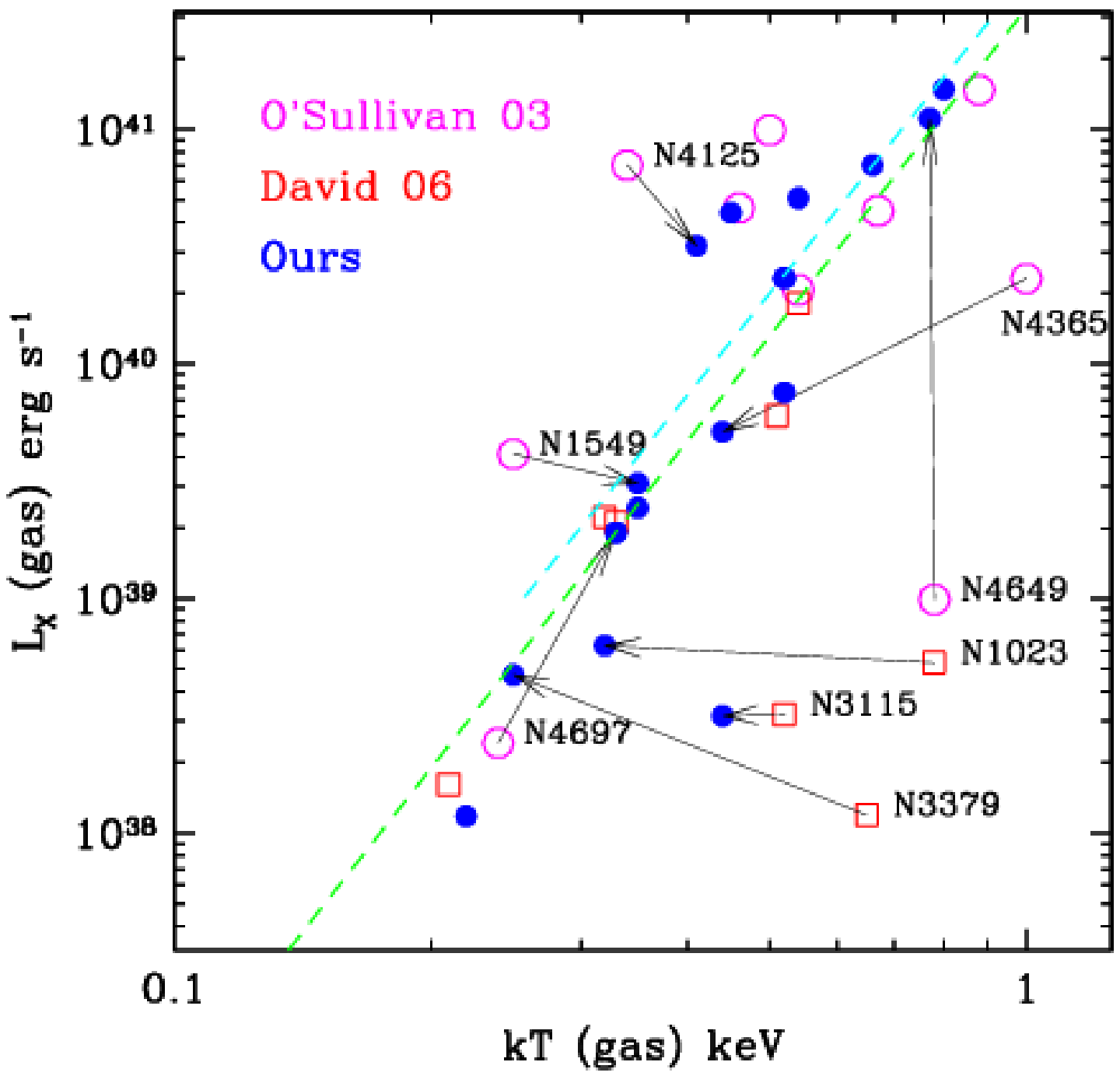
O'Sullivan 03
David 06
Ours
N4125
N4365
N1549
N4649
N1023
N3115
N4697
N3379
$L_X$ (gas) erg $s^{-1}$
kT (gas) keV
$10^{41}$
$10^{40}$
$10^{39}$
$10^{38}$
0.1
1

Figure 8

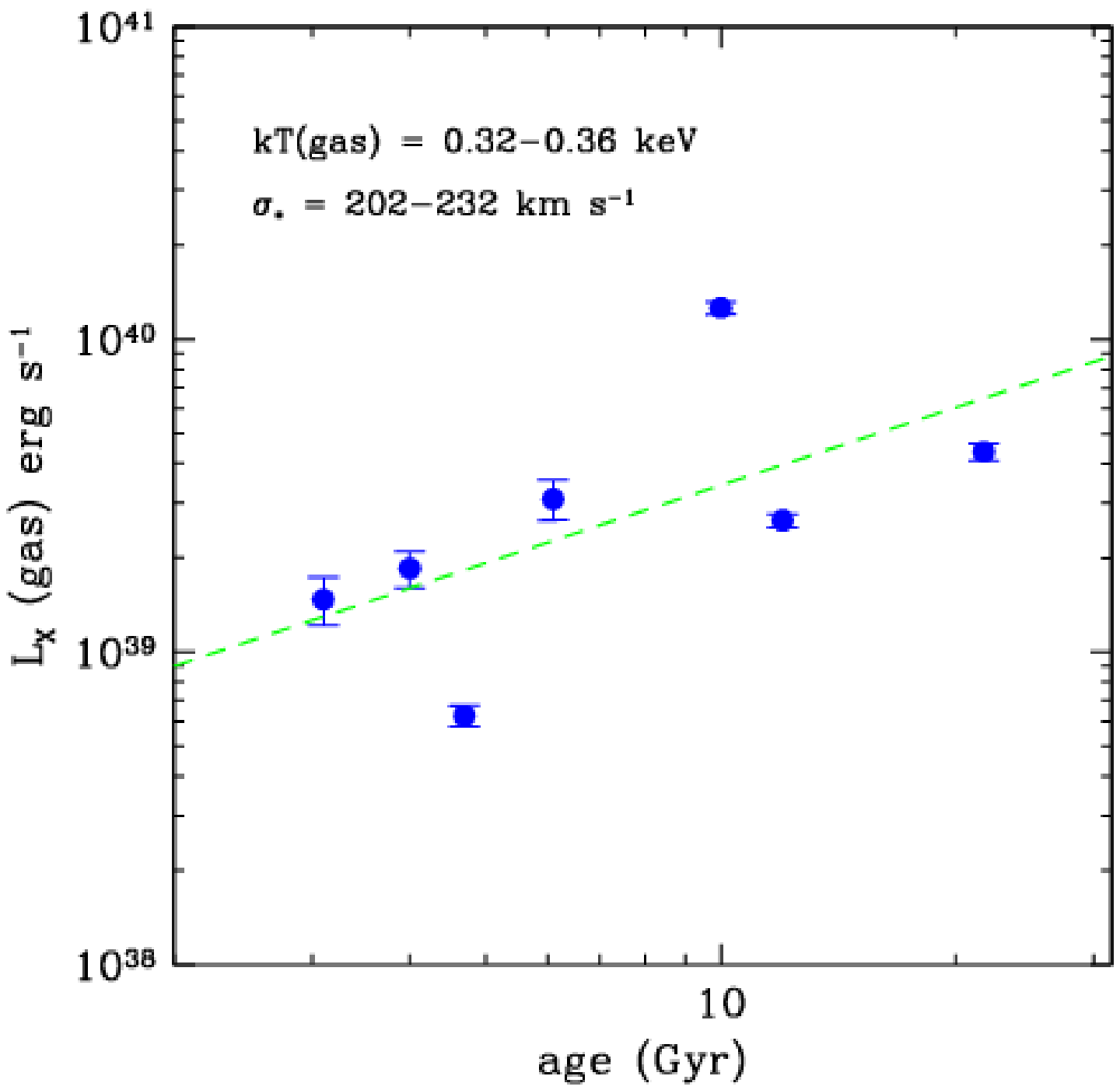
kT(gas) = 0.32–0.36 keV
σ* = 202–232 km s-1
10^41
10^40
10^39
10^38
L_X (gas) erg s-1
10
age (Gyr)

Figure 9

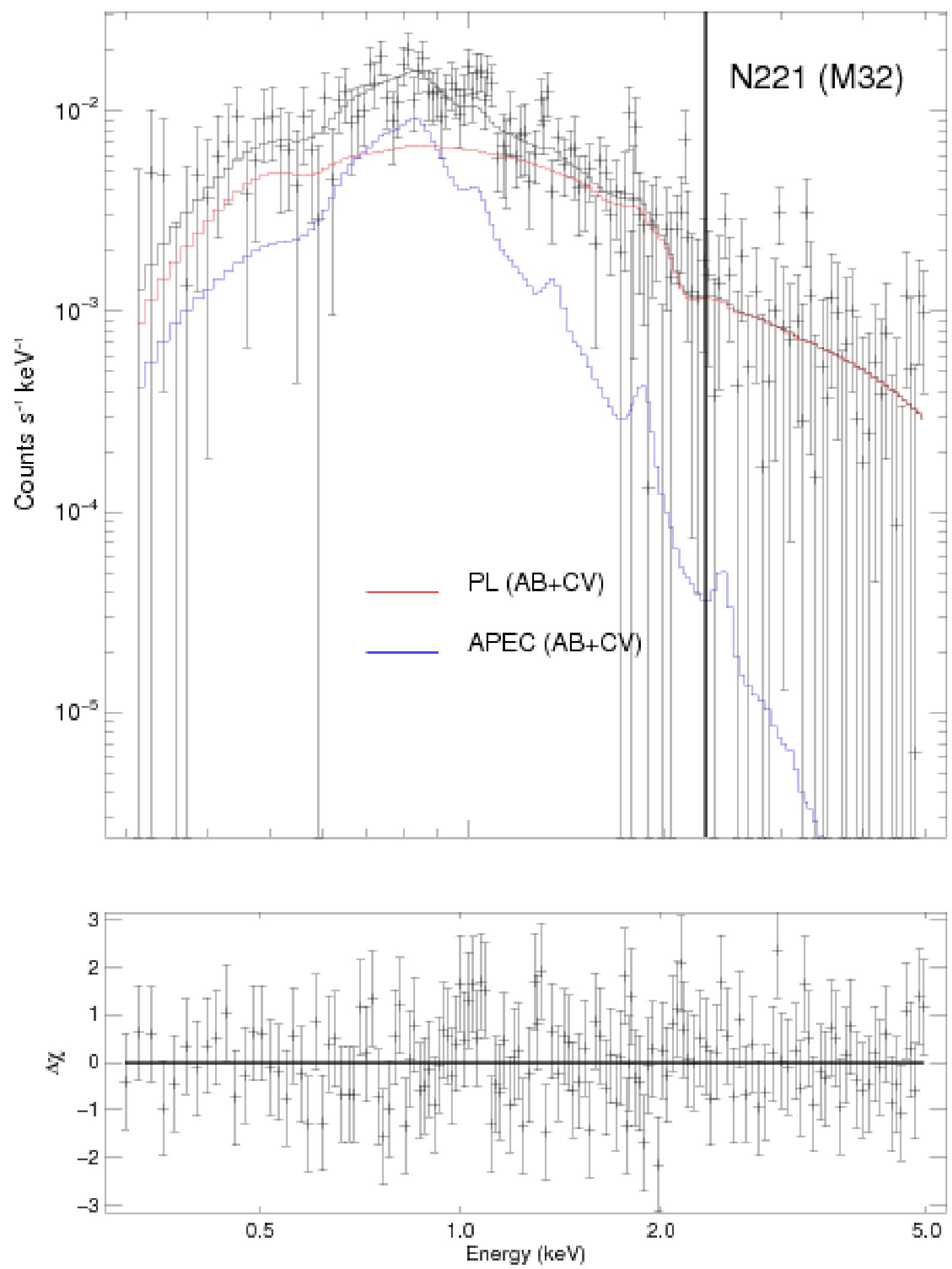
N221 (M32)
PL (AB+CV)
APEC (AB+CV)
Counts s⁻¹ keV⁻¹
10⁻²
10⁻³
10⁻⁴
10⁻⁵
Δχ
3
2
1
0
−1
−2
−3
0.5
1.0
2.0
5.0
Energy (keV)

Figure A1

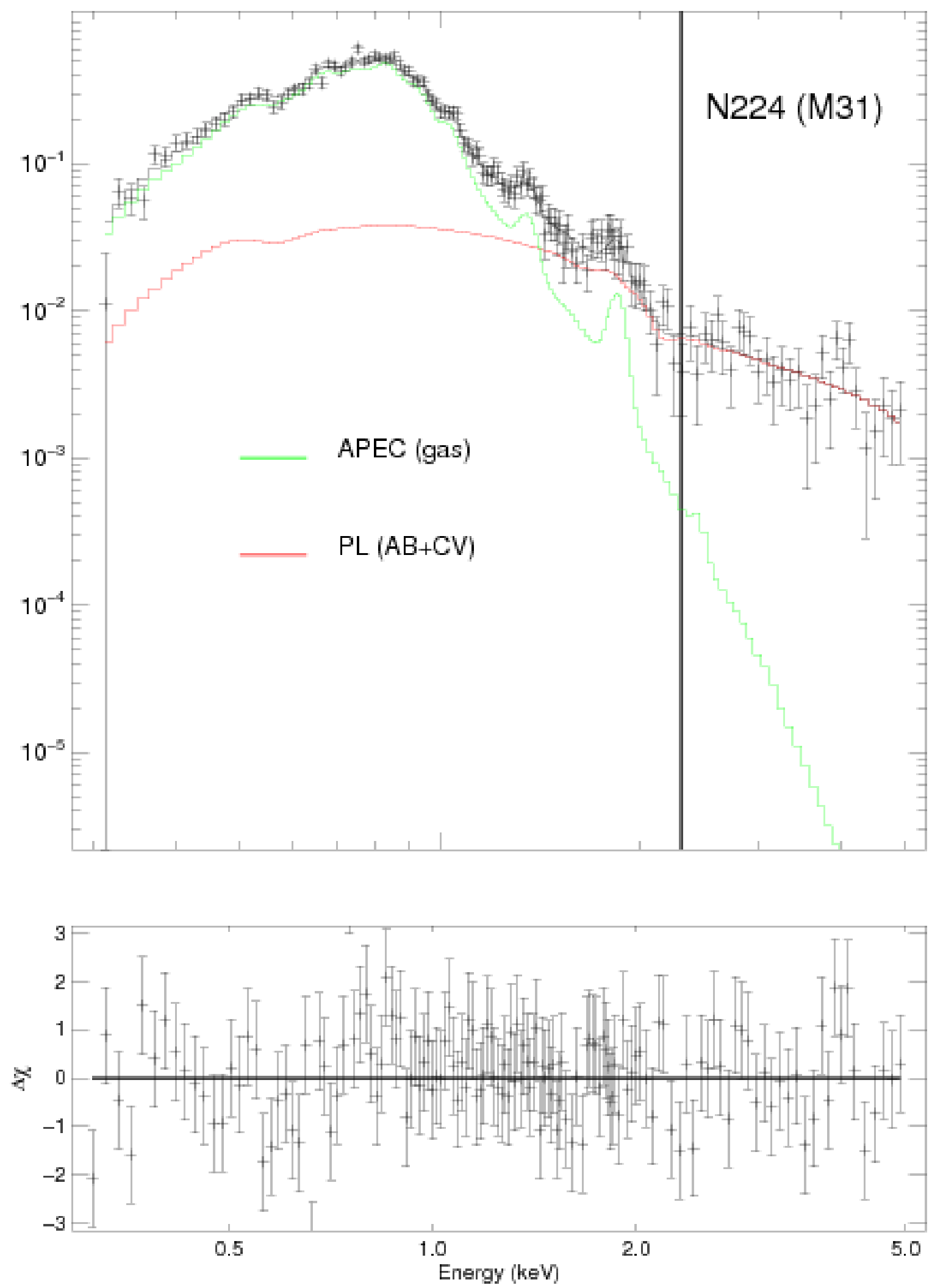

Figure A2

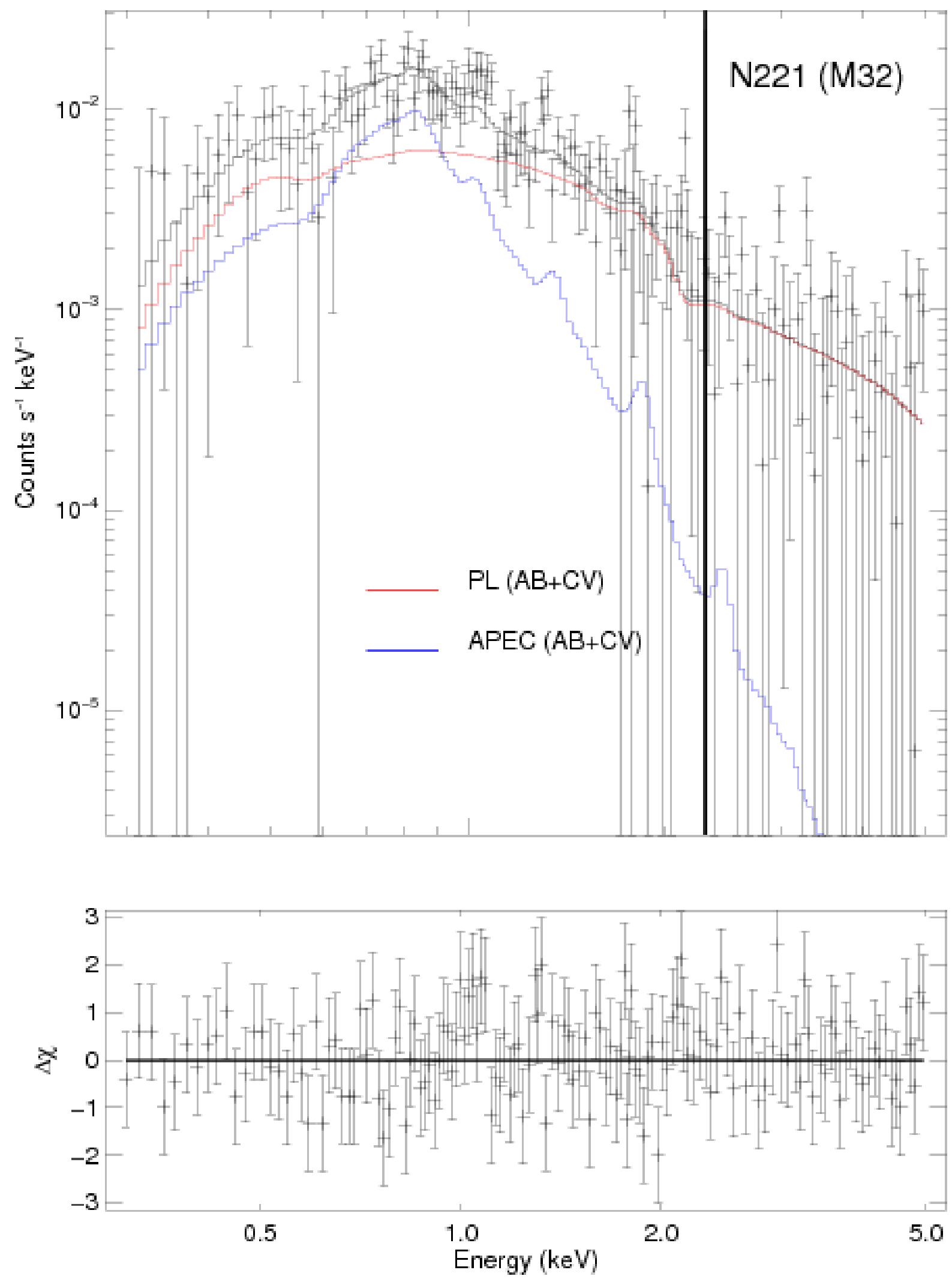
N221 (M32)
PL (AB+CV)
APEC (AB+CV)
Counts s⁻¹ keV⁻¹
10⁻²
10⁻³
10⁻⁴
10⁻⁵
Δχ
3
2
1
0
−1
−2
−3
0.5
1.0
2.0
5.0
Energy (keV)

Figure A3a

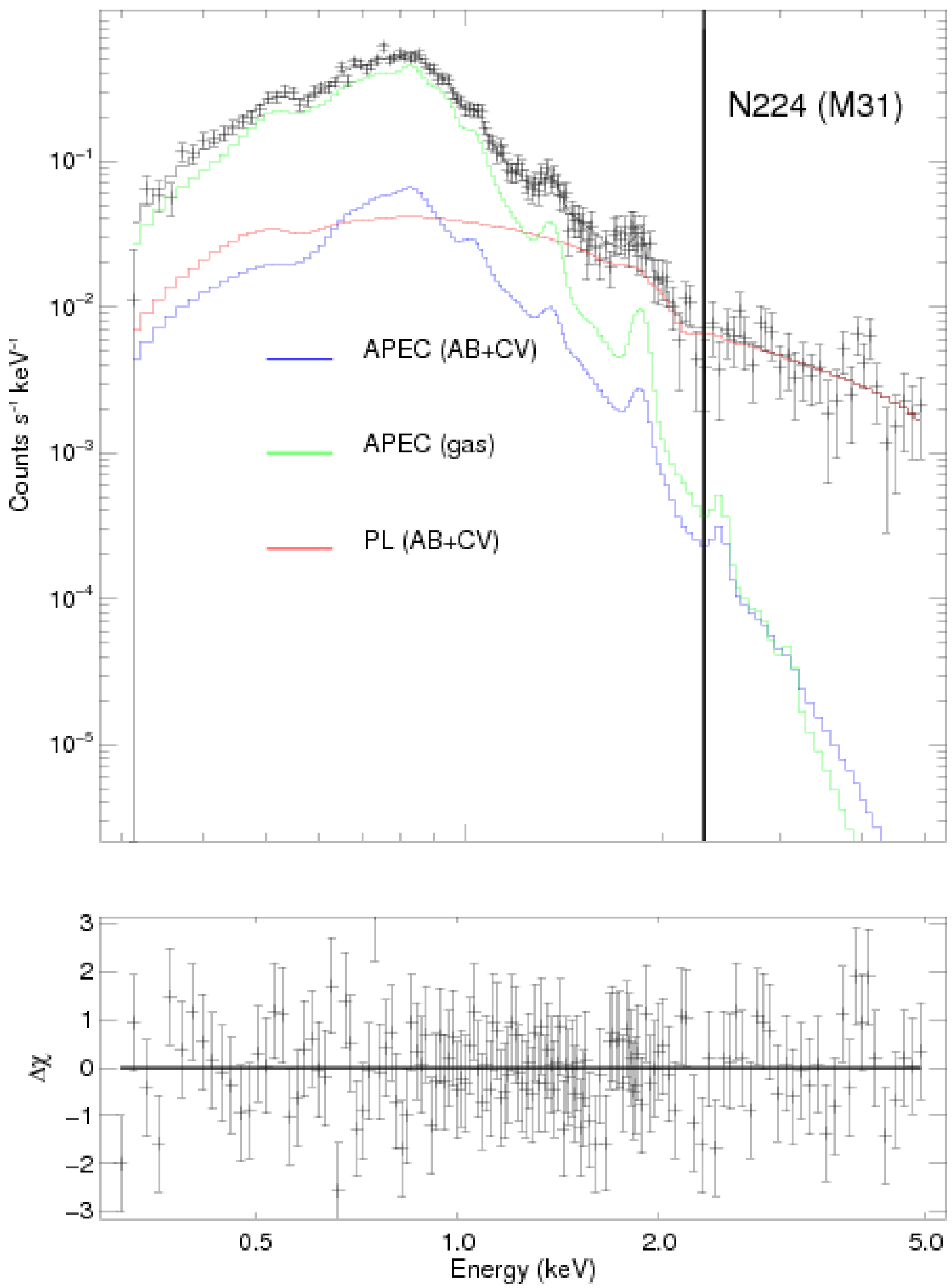

N224 (M31)
APEC (AB+CV)
APEC (gas)
PL (AB+CV)
Counts s⁻¹ keV⁻¹
10⁻¹
10⁻²
10⁻³
10⁻⁴
10⁻⁵
Δχ
3
2
1
0
−1
−2
−3
0.5
1.0
2.0
5.0
Energy (keV)

Figure A3b